\documentclass[%
 reprint,
superscriptaddress,
showpacs,preprintnumbers,
 amsmath,amssymb,
aps,
prl,
]{revtex4-1}

\usepackage{comment}
\usepackage{here}
\usepackage{txfonts}
\usepackage{amsfonts}
\usepackage[dvipdfmx]{color}%
\usepackage[dvipdfmx]{graphicx}% Include figure files
\usepackage{dcolumn}% Align table columns on decimal point
\usepackage{bm}% bold math

\begin{document}

\title{Emergence of topological semimetals in gap closing in semiconductors without inversion symmetry}
\author{Shuichi Murakami}%
\affiliation{%
 Department of Physics, Tokyo Institute of Technology, 2-12-1 Ookayama, Meguro-ku, Tokyo 152-8551, Japan
}%
\affiliation{%
 TIES, Tokyo Institute of Technology, 2-12-1 Ookayama, Meguro-ku, Tokyo 152-8551, Japan
}%
\author{Motoaki Hirayama}%
\affiliation{%
 Department of Physics, Tokyo Institute of Technology, 2-12-1 Ookayama, Meguro-ku, Tokyo 152-8551, Japan
}%
\affiliation{%
 TIES, Tokyo Institute of Technology, 2-12-1 Ookayama, Meguro-ku, Tokyo 152-8551, Japan
}%v
\author{Ryo Okugawa}
\affiliation{%
 Department of Physics, Tokyo Institute of Technology, 2-12-1 Ookayama, Meguro-ku, Tokyo 152-8551, Japan
}%

\author{Takashi Miyake}
\affiliation{%
Research Center for Computational Design of Advanced Functional Materials, AIST, Tsukuba 305-8568, Japan 
}%

\date{\today}% It is always \today, today,
             %  but any date may be explicitly specified

%\pacs{73.20.At, 74.43.Nq, 72.25.Dc}%Surface states, band structures, electron density of states,  Quantum phase transitions,  Spin polarized transport in semiconductors 
% PACS, the Physics and Astronomy
                             % Classification Scheme.
\begin{abstract}
A band gap for electronic states in crystals governs various properties of solids, such as 
transport, optical and magnetic properties. Its estimation and control have been 
an important issue in solid state physics. The band gap can be controlled externally by 
various parameters, such as pressure, atomic compositions and external field. 
Sometimes, the gap even collapses by tuning some parameter. In the field of topological insulators, such closing of the gap at a time-reversal invariant momentum indicates a
band inversion, i.e. it leads to a topological phase transition from a normal insulator 
to a topological insulator. 
Here we show that 
the gap closing in 
inversion-asymmetric crystals is universal, in the sense that the gap closing always leads either
to a Weyl semimetal or a nodal-line semimetal, from an exhaustive study on possible space groups. We here consider three-dimensional spinful systems with time-reversal symmetry. The 
space group of the system and the wavevector at the gap closing uniquely determine 
which possibility occurs and where the gap-closing points or lines lie
in the wavevector space after closing of the gap.
In particular, we show that an insulator-to-insulator transition never happens, 
which is in sharp contrast with inversion-symmetric systems. 
\end{abstract}

\maketitle
%\linenumbers

In electronic band theory of crystals, degeneracy at each wavevector ${\bf k}$ is
understood in terms of  symmetry. A dimension of an irreducible representation
of a $k$-group at a given ${\bf k}$ point is equal to degeneracy at the ${\bf k}$
point considered. For example, fourfold degeneracy of valence bands at the
$\Gamma$ point in cubic semiconductors
comes from cubic symmetry.  
Meanwhile, proposals 
of topological semimetals have shown us other possibilities for band degeneracies, 
stemming from topology. 
In  such a topological semimetal, 
a band gap closes at generic ${\bf k}$ points, and this closing of the gap  
originates not from symmetry,
but from topological reasons.
There are various topological semimetals, such as Weyl, Dirac, and nodal-line semimetals. 
In Weyl semimetals  \cite{Murakami07b, Wan,Wang12,Wang13}, the band structure have three-dimensional non-degenerate
Dirac cones. 
There are various proposals of materials for
 Weyl semimetals \cite{Wan, Xu, Liu-BiTeI, Hirayama, Weng, Huang, Rauch,Huang-SrSi2,Soluyanov-WTe2},
 some of which have been experimentally confirmed such as TaAs \cite{LvPRX, SYXua, Lv}.
Because of topological properties of Weyl nodes  \cite{Berry84, Volovik, Murakami07b}, characteristic surface states called 
Fermi  arcs arise \cite{Wan, Ojanen, Okugawa, Haldane}. 
%associated Hall effect \cite{Xu, Yang, BurkovPRL, Zyuzin}. 
%This topological properties in Weyl semimetals 
%are robust. 
%In general, 
%such robustness is not expected for the Dirac semimetals due to the 
%degeneracy between a monopole and an antimonopole. 
As another example of topological semimetals, a nodal-line semimetal \cite{BurkovPRB,Mullen15,Fang15,Chen15nano,Weng15b,Kim15,
Yu15,Xie15,Chan15,Zeng15,Yamakage16,Zhao15,Hirayama16}
has line nodes along which the band gap closes.
%In the absence of SOC, this class of semimetals is realized 
%by topological argument. 
Because emergence of topological degeneracy is accidental, 
search for candidate materials realizing topological semimetals 
is still elusive. 

\begin{figure}[htp]
\includegraphics[width=8cm]{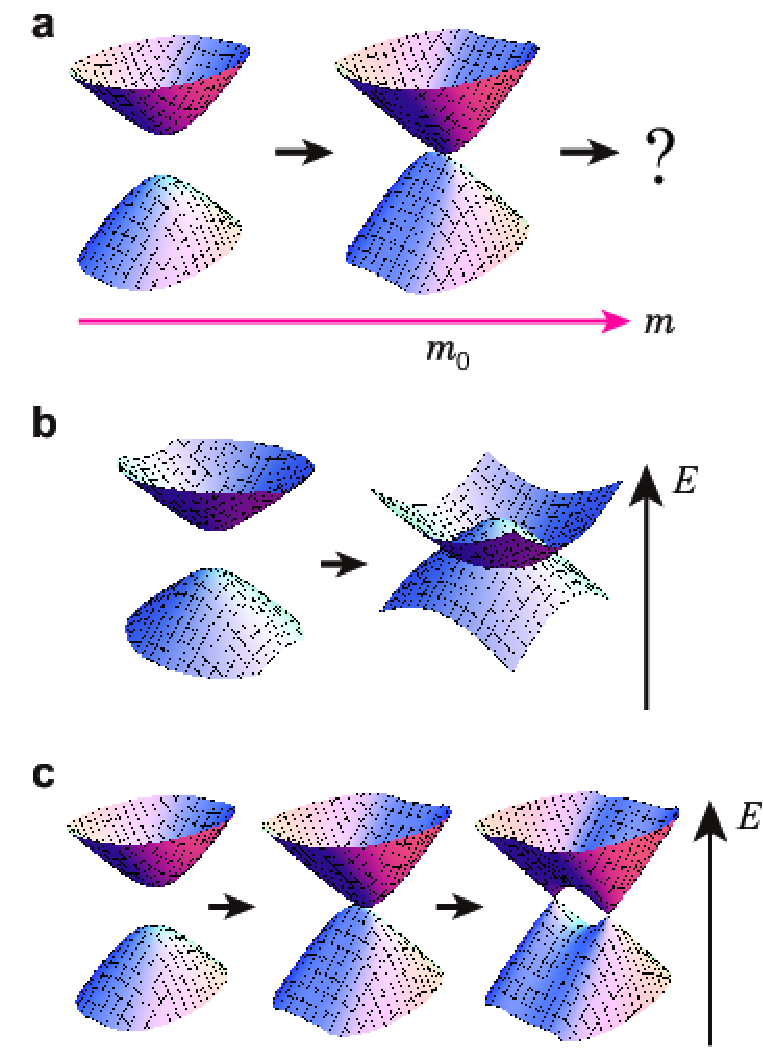}
\caption{\label{fig:sch} Setup of our main problem of band evolution in 
an inversion-asymmetric semiconductor. 
{\bf a}: Band evolution towards gap closing. 
Schematic illustrations for two classes of band evolution into {\bf b}: 
 a nodal-line semimetal, i.e. semimetal with a gap closing along a loop, and {\bf c}: a Weyl semimetals}
\end{figure}

In the present paper, we consider physics of topological semimetals from a new perspective. Here,  
we focus on evolution of a band structure of a general inversion-asymmetric insulator by 
changing a single parameter $m$, which can be any parameter in the Hamiltonian. We 
suppose that the gap closes at some value of the parameter $m$. 
We then prove that after a further change of the value of $m$  (Fig.~\ref{fig:sch} {\bf a}), the system always becomes either (A) a nodal-line semimetal (Fig.~\ref{fig:sch} {\bf b}) or 
(B) a Weyl semimetal (Fig.~\ref{fig:sch} {\bf c}).  
We show that the space group of the crystal and the wavevector at the gap-closing uniquely 
determines which possibility is realized and where the gap-closing points or lines are located
after the closing of the gap.
Here, we restrict ourselves to three-dimensional (3D)  spinful systems with 
time-reversal symmetry, i.e. nonmagnetic systems with nonzero spin-orbit coupling (SOC).
In particular, we find that an insulator-to-insulator (ITI) 
phase transition never occurs in any inversion asymmetric systems.
It is in sharp contrast with inversion-symmetric systems.

\begin{figure}[htp]
%\begin{figure*}[htp]
%\includegraphics[clip,width=0.45\textwidth,bb=0 0 1332 991]{Weyl-fig2.jpg}
\includegraphics[width=8cm]{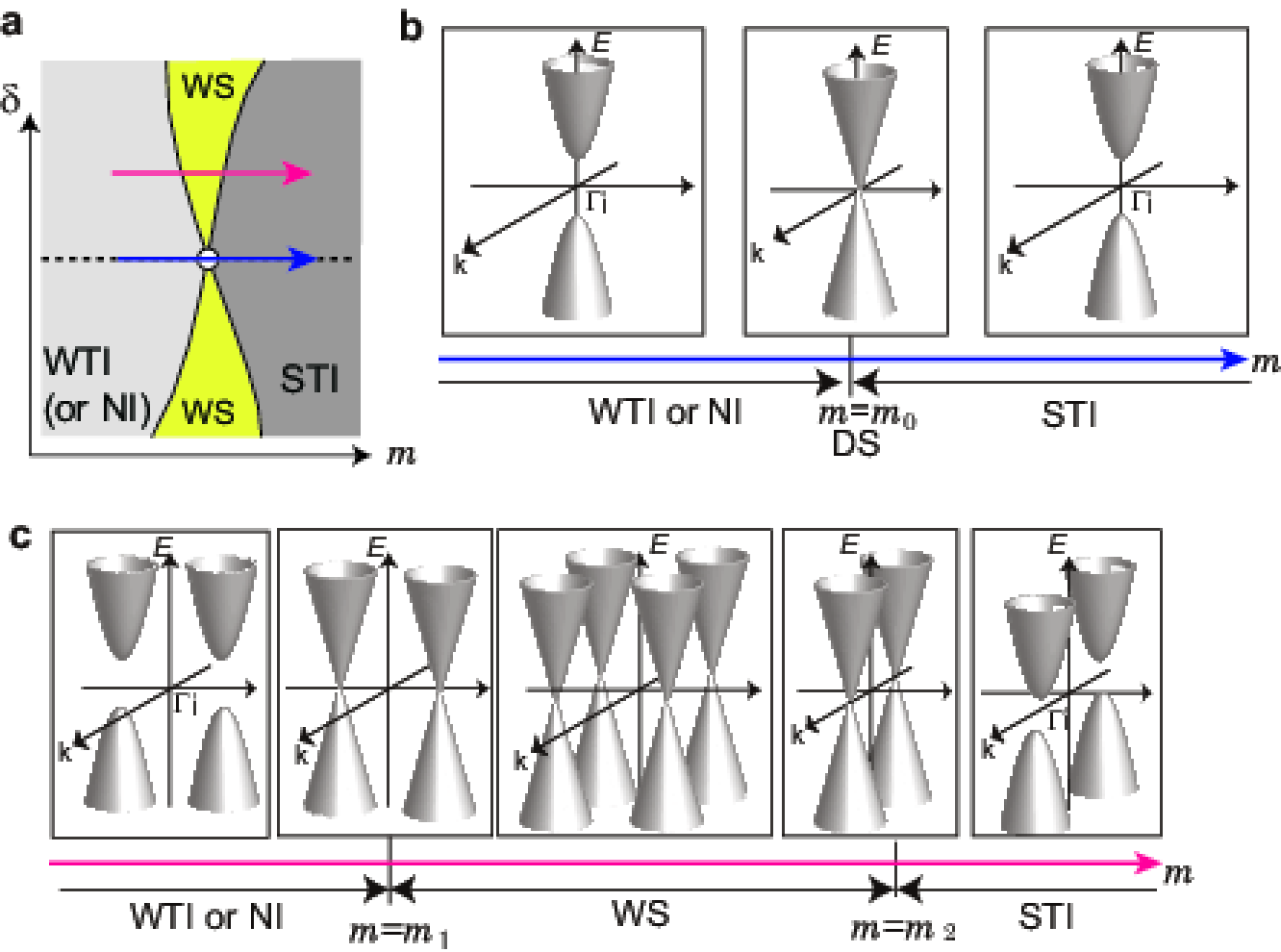}
\caption{\label{fig:fig1} {\bf a}, Universal phase diagram for $Z_2$ topological phase transitions. It is shown as a function of an external parameter $m$ and a parameter $\delta$ controlling a degree of inversion symmetry breaking \cite{Murakami07b}. The horizontal dotted line 
is the inversion-symmetric line. {\bf b}, {\bf c}, Evolutions of bulk band structure with a change of 
the parameter $m$, for inversion-symmetric ({\bf b}) and inversion-asymmetric ({\bf c})
cases, respectively. The blue and the red arrows in ${\bf a}$ corresponds to the cases {\bf b} and {\bf c}, respectively. }
%\end{figure*}
\end{figure}

This work is motivated by the universal phase diagram (Fig.~\ref{fig:fig1}{\bf a}) between a
topological insulator (TI) and a normal insulator (NI) in three dimensions \cite{Murakami07b}.
Our result in the present paper indicates that  when inversion symmetry is broken, 
at a transition between 
different $\mathbb{Z}_2$ topological numbers \cite{FKM,MooreBalents}, 
i.e. between a strong topological insulator (STI) and 
a normal insulator (or a weak topological insulator (WTI)),
a Weyl semimetal (WSM) phase always appears, as is expected from Ref.~\cite{Murakami07b}. 
This theory can be applied to any inversion asymmetric crystals, such as BiTeI under high pressure \cite{Liu-BiTeI} and Te under pressure \cite{Hirayama16}.

{\bf Setup of the problem}--
We consider a Hamiltonian matrix  $H({\bf k},m)$ where ${\bf k}$ is
a Bloch wavevector, and $m$ is an external parameter controlling the gap. 
Furthermore, we assume that 
a space group of the system remains the same for any values of $m$. 
To see how the gap closes, we assume that for $m<m_0$ the 
system is an insulator, and that at $m=m_0$ the gap closes at a wavevector ${\bf k}={\bf k}_0$.
We assume the Hamiltonian to be analytic with respect to $\bf{k}$ and $m$.
We then expand the Hamiltonian in terms of $m-m_0$ and ${\bf q}\equiv{\bf k}-{\bf k}_0$, and 
retain some terms of lower order, in order to see evolution of the band structure 
for $m>m_0$. We consider 
all the 138 space groups without inversion symmetry, for the purpose of application to real materials. 
For each space group, there are various ${\bf k}$ points such as $\Gamma$, $X$ and $L$.
Each ${\bf k}$ point is associated with a $k$-group (little group), which leaves the ${\bf k}$ point unchanged. 
In order to focus on the closing of the gap, we retain only the 
lowest conduction band and the highest valence band, 
whose irreducible representations (irreps) of the $k$-group at ${\bf k}_0$ are denoted as $R_{\rm v}$ and $R_{\rm c}$, respectively. 
In our analysis, we use the complete list of 
double-valued irreps of $k$-groups  in Ref.~\cite{Bradley}. 

\begin{figure}[htp]
\includegraphics[width=8.5cm]{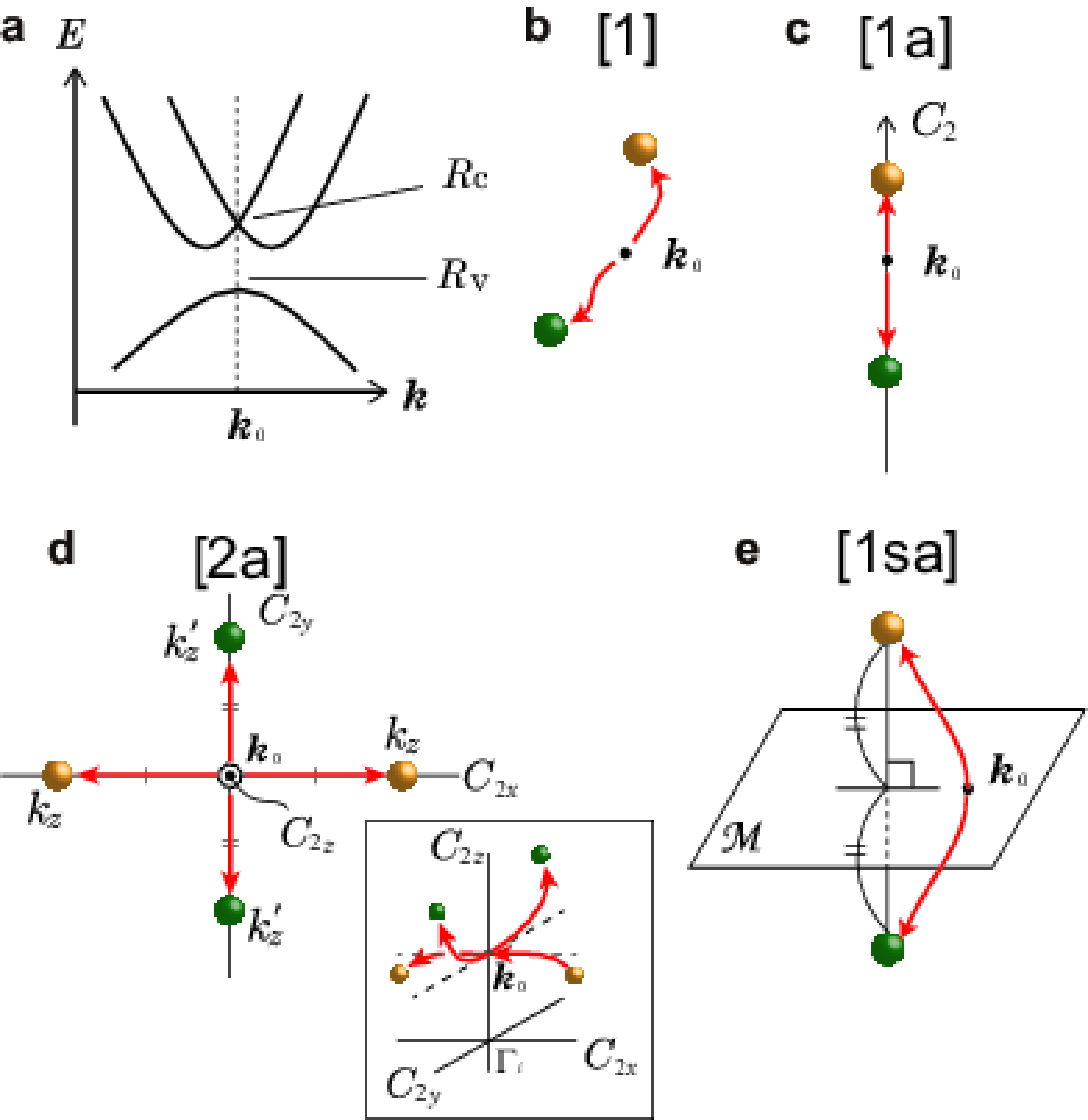}
\caption{\label{fig:m0}{\bf a}, Schematic band structure for the  case 
with ${\rm dim}R_{\rm c}=2$ and ${\rm dim}R_{\rm v}=1$.  The gap 
does not close at ${\bf k}_0$ in this case.
{\bf b}, Trajectory of Weyl nodes in (i) after pair creation at ${\bf k}_0$.
{\bf c}, Trajectory of Weyl nodes in (ii-2) and in (iii-2). The Weyl nodes move along the 
$C_2$ axis. {\bf d}, Trajectory of Weyl nodes in the case (iii-1), viewed from different angles. The ${\bf k}_0$ point
is on the $C_{2z}$ axis away from a TRIM $\Gamma_i$.
{\bf e}, Trajectory of Weyl nodes in the case (iv-1).  In 
{\bf b}-{\bf e}, yellow and green spheres denote
monopoles and antimonopoles in ${\bf k}$ space, respectively, and they are both Weyl nodes. }
\end{figure}

Varieties of 138 space groups, ${\bf k}$ points for each space group, and irreps at each ${\bf k}$ point 
lead to numerous possibilities. 
Our theory in this paper exhausts all the cases; in fact, we can substantially reduce the number of possible cases 
by the following considerations. First, we can exclude cases with 
${\rm dim}R_{\rm v}\geq 2$ or ${\rm dim}R_{\rm c}\geq 2$, i.e. the cases where
the valence or the conduction band at ${\bf k}_0$ have degeneracy; it is 
because the gap does not close at ${\bf k}_0$. 
%Here the dimensions of the irreps includes the
%degeneracy due to the time-reversal symmetry.
The reason is the following. 
If ${\rm dim}R_{\rm c}\geq 2$ at ${\bf k}_0$, the energy of the conduction band 
does not have a minimum at 
this ${\bf k}_0$ point, because the degenerate states at ${\bf k}_0$ 
will split linearly away from ${\bf k}_0$ along a generic direction (see Fig.~\ref{fig:m0}{\bf a}).
Therefore the gap does not close at ${\bf k}_0$. A similar argument applies to 
the valence band. 
Thus 
we can safely restrict ourselves to the cases with ${\rm dim}R_{\rm v}={\rm dim}R_{\rm c}=1$; 
hence, the effective model is described by 
a $2\times 2$ Hamiltonian matrix, which drastically simplifies our analysis. 
In particular, among possible wavevectors ${\bf k}_0$, we exclude the time-reversal invariant momenta (TRIM),  
because the bands  at the TRIM always have Kramers degeneracies.

{\bf Various cases of gap-closing events classified by symmetry}-- 
Next, we determine the Hamiltonian matrix  $H({\bf k},m)$ for each case from  $R_{\rm v}$ and $R_{\rm c}$.
Resulting behaviors of  the closing of the gap are then
classified in term of $R_{\rm v}$ and $R_{\rm c}$. 
All the cases for 138 space groups are summarized in the tables in the Supplemental Material. 
Here, we briefly explain some representative cases of $k$-groups, and their details are given in the Methods section. 
In the analysis here, in addition to point-group elements of a $k$-group, we have to consider symmetry operations of a form $\Theta O$, where 
$\Theta$ is a time-reversal operation and $O$ is a point-group element.
. 

{\bf (i) no symmetry}: 
We consider a generic ${\bf k}$ point having no special symmetry; namely, a $k$-group 
consists only of an identity operation. The band gap can close there, and the closing of the gap 
always accompanies a pair creation of Weyl nodes, as shown in Ref.~\cite{Murakami07b,MurakamiKuga}. 
This occurs because Weyl nodes are topological, having quantized monopole charges $q=\pm 1$ for the
${\bf k}$-space Berry curvature; this topological property allows pair creation of Weyl nodes 
with $q=+1$ and $q=-1$.
We call this case of Weyl-node creation as ``{\bf 1}'', where {\bf 1} represents a number of 
monopole-antimonopole pairs (Fig.~\ref{fig:m0}{\bf b}).

{\bf (ii) $C_{2}$ symmetry}: Suppose the ${\bf k}$-group consists only of the twofold ($C_2$) symmetry. 
Then there are two one-dimensional (1D) 
irreps with opposite signs of the $C_2$ eigenvalues. 
Consequently we have two cases: (ii-1) $R_{\rm c}=R_{\rm v}$ and 
(ii-2) $R_{\rm c}\neq R_{\rm v}$.
For (ii-1): $R_{\rm c}=R_{\rm v}$, the gap cannot close by changing $m$, because of level repulsion.
On the other hand, for (ii-2): $R_{\rm c}\neq R_{\rm v}$, the gap can close because there is
no level repulsion at ${\bf k}_0$, and 
we find that the closing of the gap 
accompanies creation of a pair of Weyl nodes. 
When $m$ is increased, the two Weyl nodes (a monopole and an antimonopole) move along the 
$C_2$ axis, as shown in Fig.~\ref{fig:m0}{\bf c}.
We call this case ``{\bf 1a}'', where {\bf 1} represents the number of 
monopole-antimonopole pairs, and ${\bf a}$ means ``axial'',  i.e. the relative direction 
between the two Weyl nodes is along a high-symmetry axis. 

(iii) {\bf $C_2$ and $\Theta C_2$ symmetries}: 
The problem becomes more complicated
when the time-reversal operator $\Theta$ is involved. 
Here, we consider a system with two $C_2$ symmetries $C_{2y}$, $C_{2z}$,
whose rotational axes ($y$ and $z)$ cross perpendicularly.
We focus on a ${\bf k}_0$ point lying on the $C_{2z}$ axis, but not on the $C_{2y}$ axis; 
the $k$-group then consists of $C_{2z}$ and $\Theta C_{2y}$. 
There are two eigenvalues of $C_{2z}$ with opposite signs, yielding two irreps, 
and $R_{\rm c}$ and $R_{\rm v}$ can take either of these two irreps. 
We consider two cases (iii-1) $R_{\rm c}=R_{\rm v}$, and (iii-2) $R_{\rm c}\neq R_{\rm v}$ separately.
For (iii-1) $R_{\rm c}=R_{\rm v}$, the closing of the gap leads to
creations of two monopoles
and two antimonopoles, and their trajectories are shown in 
Fig.~\ref{fig:m0}{\bf d}. 
As compared to (ii-1), 
the additional 
$\Theta C_{2y}$ symmetry  suppresses level repulsion at ${\bf k}_0$, and 
therefore the gap can close.
%One can close the gap, but because of the level repulsion
% the gap immediately opens
%on the $C_2$ axis 
In our list this behavior is called
${\bf 2a}$. Here {\bf 2} means the number of monopole-antimonopole pairs, and ${\bf a}$ means
that the relative orientation between the two monopoles (and likewise the two antimonopoles) 
is fixed to be along high-symmetry axes.
For (iii-2)  $R_{\rm c}\neq R_{\rm v}$, 
a pair of Weyl nodes are created and they 
move along the $C_{2z}$ axis 
as in (ii-2) (Fig.~\ref{fig:m0}{\bf c}). It is written as ${\bf 1a}$. 

{\bf (iv) mirror symmetry ${\cal M}$}:  We consider the $k$-group
consisting only of a mirror symmetry or a glide symmetry. 
There are two representations with opposite signs of mirror (or glide) eigenvalues. 
In the case (iv-1): $R_{\rm c}=R_{\rm v}$, the gap can close, leading to 
a creation of a pair of Weyl nodes. The trajectory of the monopole and that of 
the antimonopole are mirror images to each other. We call this pattern {\bf 1sa}, 
representing that the direction between the two Weyl nodes are always 
perpendicular to the mirror plane (Fig.~\ref{fig:m0}{\bf e}). The symbol {\bf s} means 
that all the Weyl nodes are related by symmetry operations, and therefore they are 
at the same energy. 
On the other hand, for (iv-2): $R_{\rm c}\neq R_{\rm v}$, 
the gap closes along a loop (i.e. nodal line) within the mirror plane in ${\bf k}$ space. We call this case ${\bf 1\bm{\ell}}$, 
where ${\bf \bm{\ell}}$ stands for ``loop'' and ${\bf 1}$ represents the number of loops.  
This degeneracy along the loop occurs because on the mirror plane there is no level repulsion
between the valence and the conduction bands, belonging to the different irreps.

{\bf 138 inversion-asymmetric space groups}-- 
We can similarly calculate band-structure evolution after closing of the gap  for
all the high-symmetry points and lines for the 138 space groups without inversion symmetry. A complete list of all the cases
is lengthy and is summarized in the tables in the Supplemental Material. 
In Fig.~\ref{fig:3}, we  summarize possible patterns for positions where the gap closes in 
${\bf k}$ space for $m>m_0$, i.e. after closing of the gap at ${\bf k}_0$. 
In the figure, the individual patterns are 
represented as ${\bf 1sp}$, ${\bf 1\bm{\ell}}$, and so on,
and their notations are explained in the caption.

\begin{figure*}[htp]
\includegraphics[width=15cm]{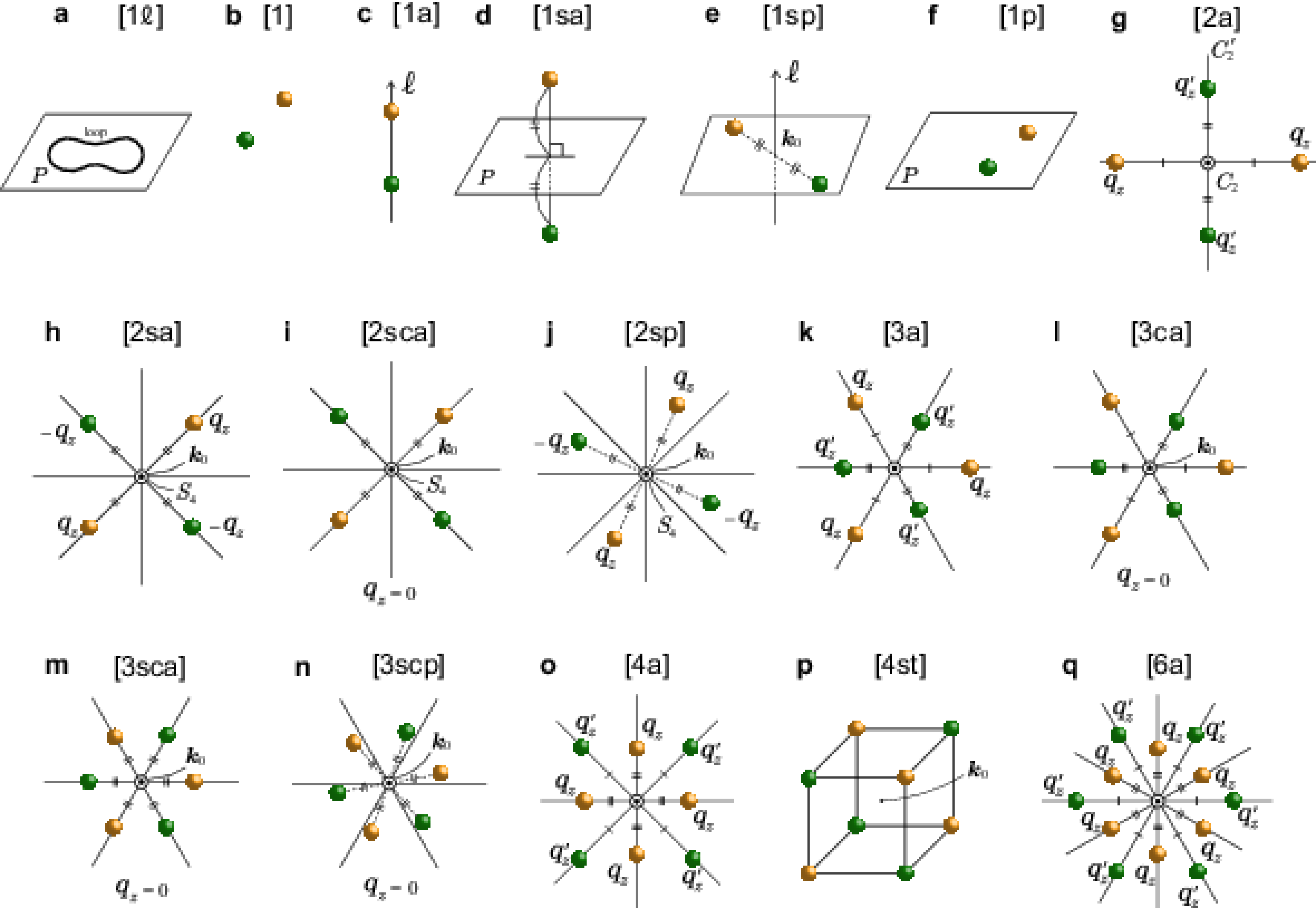}
\caption{\label{fig:3} All the patterns of locations of gap-closing points after parametric gap closing. 
{\bf a}, A line node for a nodal-line semimetal. 
While the figure shows the case with a single line node, the number of line nodes can range among 1, 2, 3, 4 and 6. {\bf b}-{\bf p}, All the patterns of Weyl nodes after parametric gap closing at ${\bf k}_0$. Yellow and green spheres denote
monopoles and antimonopoles in ${\bf k}$ space, and they are both Weyl nodes. 
Solid lines 
denote high-symmetry directions. In {\bf g-o} and {\bf q}, 
The $z$ axis, taken to be perpendicular to the page,
is parallel to a high-symmetry axis. 
${\bf q}\equiv {\bf k}-{\bf k}_0$ is the 
momentum measured from ${\bf k}_0$. In {\bf i}, {\bf l}, {\bf m} and {\bf n}, all the 
Weyl nodes are coplanar with ${\bf k}_0$, while in {\bf g}, {\bf h}, {\bf j}, {\bf k}, {\bf o}, and 
{\bf q}, the Weyl nodes are mutually displaced along the $z$ direction.
The numbers in the figures {\bf b}-{\bf q} represent the number of pairs of Weyl nodes, except for ${\bf 1\bm{\ell}, 2\bm{\ell},3\bm{\ell}, 4\bm{\ell}}$, and ${\bf 6\bm{\ell}}$.
where the number represents the number of loops. The symbol ${\bf \bm{\ell}}$ means that there 
are nodal lines where the gap is closed. Such nodal lines always appear on mirror planes, 
and only when the valence and the conduction bands have different 
mirror eigenvalues. 
The symbol ${\bf a}$ (axial) 
represents that the relative directions between the Weyl nodes are fixed to be along certain high-symmetry 
lines; meanwhile, the symbol ${\bf p}$ (planar) represents that such directions are not confined along
high-symmetry axes, but confined within high-symmetry planes. 
The symbol ${\bf c}$ (coplanar) means that all the monopoles and antimonopoles 
lie on a same high-symmetry plane. This symbol ${\bf c}$ is used only when there are more than one 
monopole-antimonopole pairs. The symbol ${\bf t}$ (tetrahedron) appears only for 
a few cases with tetrahedral or cubic symmetries. It means that 
4 monopoles and 4 antimonopoles form 8 vertices of a cube whose center is a high 
symmetry point, and 4 monopoles form a tetrahedron. The symbol ${\bf s}$ (symmetric) means that all the monopoles and the antimonopoles are related to each other 
by symmetry operations. In such cases
they are energetically degenerate, and it is possible to locate all the Weyl nodes 
on the Fermi energy. Otherwise, these monopoles and antimonopoles may not necessarily be at
the same energy. }
\end{figure*}

From this analysis, we conclude that there are only two possibilities after closing of the gap 
in inversion-asymmetric insulators with time-reversal symmetry: nodal-line semimetals and 
Weyl semimetals. 
The nodal-line semimetals are denoted as $n{\bf \bm{\ell}}$  
$(n=1,2,3,4,6)$ (See Fig.~\ref{fig:3} {\bf a}). Here $n$ is the number of nodal-lines on mirror planes, on which the highest valence and the lowest conduction 
bands have different mirror eigenvalues.
The other possibility is the Weyl semimetal, 
shown in Fig.~\ref{fig:3}{\bf b}-{\bf q}.
%The gap closing corresponds to monopole-antimonopole pair creations of Weyl nodes. Because of the topological nature of 
%Weyl nodes (monopoles or antimonopoles), these Weyl nodes move without
%changing its monopole charge. 
In some ${\bf k}$ with high symmetry, Weyl nodes and nodal lines are simultaneously generated. 
Remarkablely, an ITI transition never occurs in closing of the gap of the inversion-asymmetric
insulators.

{\bf Materials realization}--
This universal result applies to any crystalline materials without inversion symmetry,
as we show some examples in the following.
Our first example is 
 tellurium (Te), which has been theoretically
shown to become a WSM at high pressure \cite{Hirayama}. 
Tellurium is a narrow gap semiconductor
without inversion symmetry, with its space group No.152 ($P3_121$) or No.154  ($P3_221$), which are mirror images to 
each other. 
At higher pressure, the gap closes and eventually a pair of Weyl nodes
is produced at each of the four P points on the K-H lines. The Weyl nodes then move along
the  $C_3$ axes (K-H lines). 
%Because 
%this K-H lines are $C_3$ axes, and the gap closes between 
According to our table, the only possibility of the gap closing is ${\bf 1a}$, 
in agreement with the above result. Moreover, this 
${\bf 1a}$ at the P points
is allowed only when the $C_3$ eigenvalues of the valence and conduction bands 
are different; it was confirmed by the 
{\it ab initio} calculation \cite{Hirayama}.

The second example is HgTe$_{x}$S$_{1-x}$ under strain, which has been
shown to become a WSM \cite{Rauch}.
It has zinc-blende structure (space group No216, $F\bar{4} 3 m$), but a 
strain along the [001] direction reduces the space group to No.119 ($I\bar{4}m2$).
%HgS and HgTe show STI phases under positive in-plane strain. %to lift degeneracy at $\Gamma$.
In this case, when $x$ is increased from $x=0$, the gap closes at four points on the $\Gamma$-K lines ($\Sigma$ points)
on the (110) and the (1$\bar{1}$0) mirror planes, and four pairs of Weyl nodes are created. The eight Weyl points then move within the $k_z=0$ plane by a further increase of $x$, until they 
mutually annihilate at $\Sigma$ points after $\pm\pi/2$ rotation around the 
$[001]$ axis \cite{Rauch}. 
According to our table, the gap closing at each $\Sigma$ point corresponds to ${\bf 1sa}$, meaning that the 
Weyl nodes move perpendicular to the (110) or the (1$\bar{1}$0) mirror planes. It agrees
with the previous work \cite{Rauch}.

\begin{figure*}[htp]
\includegraphics[width=15cm]{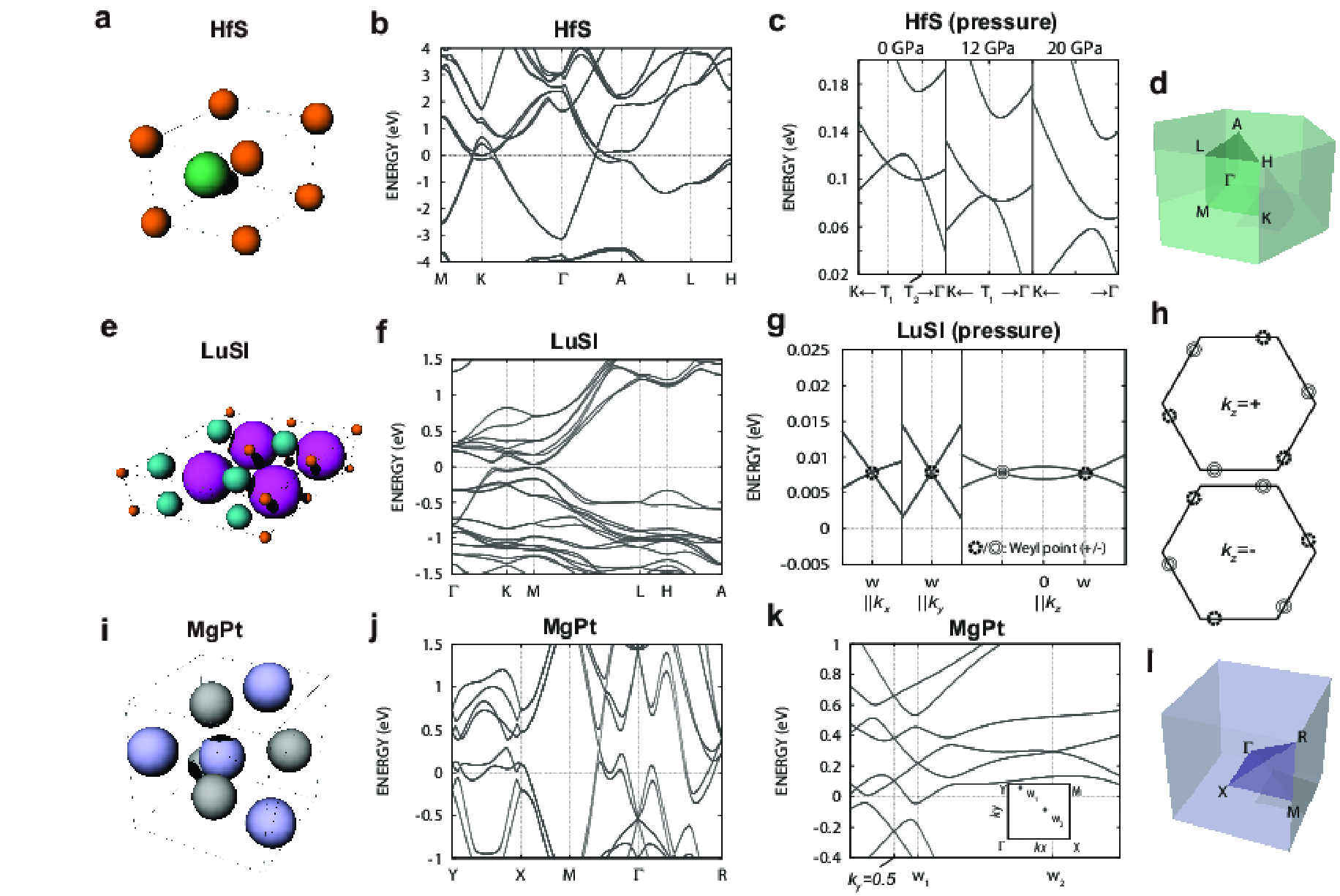}
\caption{\label{fig:mat} {\bf a}, Crystal structure of HfS. {\bf b}, {\bf c},
electronic band structure of HfS.
{\bf c} shows the phase transition of HfS under pressure.
T$_1$ and T$_2$ are the points of the intersection of the nodal-line with K$\Gamma$ line.
{\bf d}, Brillouin zone of HfS.
{\bf e}, Crystal structure of LuSI. 
{\bf f}, {\bf g},
Electronic band structure of LuSI.
{\bf g} Band structure of 
LuSI under pressure, where lattice constant $c$ is multiplied by $0.945$.
{\bf h} Positions of the Weyl nodes of LuSI under pressure on the $k_z = 0$ plane.
{\bf i}, Crystal structure of MgPt. 
{\bf j}, {\bf k},
electronic band structure of MgPt.
{\bf l}, the Brillouin zone of LuSI and MgPt.
The energy is measured from the Fermi level.}
\end{figure*}

Because degeneracies in topological semimetals are accidental, 
efficient and systematic search of  topological semimetals is difficult. 
Moreover, such degeneracies occur at generic ${\bf k}$ points;
therefore, they may easily be overlooked in {\it ab initio} calculations, where 
band structure is usually calculated only along high-symmetry lines.
Thus, a search of topological semimetals has been an elusive issue. 

Our results can be used for search of topological semimetal materials.
For example, we find that HfS has nodal lines near the Fermi level.
HfS has a valley of the density of states near the Fermi level  because of the following two reasons.
One is 
covalent bonds from S $3p$ orbitals, by which valence electrons in $p$ orbitals tend to make a gap at the Fermi level to lower the total energy; it is prominent for atoms with large electronegativity.
The other is an open shell of  Hf $5d$ orbital ($5d^2$)
%both the bottom of the conduction band and the top of the valence band originate from the
% Hf $5d$ orbitals 
having strong SOC.
%Next, we discuss the effect of the structure and its symmetry. 
The structure of HfS is the same as that of tungsten carbide (WC) (Fig.~\ref{fig:mat} {\bf a}), with the space group No.187 ($P\bar{6}m2$)~\cite{Franzen}.
Figure~\ref{fig:mat} {\bf b} shows electronic band structure of HfS.
If the SOC is neglected, Dirac nodal lines exist around the K points on the $k_z =0$ plane.
The SOC lifts the degeneracy of the Dirac nodal lines, and Weyl nodal lines appear instead, near the K points on the mirror plane 
$k_z =0$.
%T$_1$ and T$_2$ are the points of the intersection of the nodal-line with K$\Gamma$ line.
%The band dispersion at the nodal-line is parabolic for the $\Gamma $M direction.
%The nodal-line is on the $k_z =0$ plane with ${\cal M}$, which is consistent with our discussion.
By applying pressure (Fig.~\ref{fig:mat} {\bf c}) or by atomic substitution from S to Se, the nodal lines become smaller.
%Here, we show the annihilation of the nodal-line by changing pressure .
At 9 GPa, the nodal lines shrink to points (T points) on
the K$\Gamma$ lines, and 
then the gap opens above 9 GPa.
This evolution of the band structure corresponds to ${\bf 1 \bm{\ell}}$ in our table.

Next, we show  LuSI as a Weyl semimetal under pressure.
The conduction and the valence bands originate from Lu $6s+5d$ orbitals, and S $3p$ (and I $5p$) orbitals, respectively.
The structure type of LuSI is GdSI (Fig.~\ref{fig:mat} {\bf e}), with the space group No.174 ($P\bar{6}$)~\cite{Dagron}.
Each type of atoms constitutes a distorted trigonal lattice,
which displaces the gap away from the $\Gamma $ point.
%This space group has a possibility of ${\bf 3scp}$ type gap-closing in Fig.~\ref{fig:3} ${\bf m}$,
%where all of the Weyl point are related to each other by symmetry and thus have same energy.
Figure \ref{fig:mat} {\bf f} shows electronic structure of LuSI, having a 
very narrow gap ($< 0.04$ eV) near the M point.
It becomes a Weyl semimetal under pressure, as shown in Fig.~\ref{fig:mat} {\bf g}~\cite{P_LuSI}.
There exist 6 monopoles and  6 antimonopoles at the same energy because 
they are related by symmetry.
By applying pressure, the band gap first closes at six generic points on the $k_z=0$ plane, related with each other by 
sixfold symmetry. 
It corresponds to {\bf 1sa} because of 
the mirror symmetry ${\cal M}_z$, belonging to the case (iv-1).

We show another example, MgPt, as a material having Weyl nodes. 
MgPt has the FeSi-type structure (Fig.~\ref{fig:mat} {\bf i}) having the space group No.198 ($P2_13$)~\cite{Stadelmaier}.
%The FeSi-type structure could be regarded as the distorted NaCl-type structure  ($Fm$-$3m$), where the atom is displaced along the [111] axis. 
%The gap-closing occurs only at the Weyl-node in this system, 
%because the space group does not have mirror inversion plane.
Figures~\ref{fig:mat} {\bf j, k} show electronic structure of MgPt.
The band near the Fermi level originates from Pt $5d$ orbitals having strong SOC.
Weyl nodes w$_1$ and w$_2$ exist at general $\bm{k}$ points with no symmetry.
%In MgPt, the distance between the Weyl nodes is quite long, 
%is much longer than other Weyl semimetals proposed at present,
%which would give large Fermi arcs on the surface.

{\bf Topological phase transitions and $Z_2$ topological number}--
Let us turn to $\mathbb{Z}_2$ phase transitions in three dimensions, i.e. 
STI-NI (or STI-WTI) phase transitions. In 
the universal phase diagram (Fig.~\ref{fig:fig1} {\bf a})
between a STI and a NI (or a WTI),
there should be a finite region of a Weyl semimetal phase when inversion symmetry is broken as shown in the previous works \cite{Murakami07b,MurakamiKuga};
nevertheless in its derivation \cite{Murakami07b},
crystallographic symmetries except for inversion symmetry are not considered.
The result in the present paper shows that this conclusion of existence of the Weyl semimetal phase in the $Z_2$ phase transition
holds true in general, whenever inversion symmetry is broken. 
One of the remarkable 
conclusions here is that an ITI transition never occurs in inversion-asymmetric crystals. 
This is in strong contrast with inversion-symmetric systems,
where a transition between different $Z_2$ topological 
phases always occurs as an ITI transition (i.e. at a single value of $m$), as 
seen in 
TlBi(S$_{1-x}$Se$_x$)$_2$ 
at around $x\sim 0.5$ \cite{Xu-TlBiSSe,Sato-TlBiSSe}. 
%
%Monopoles and antimonopoles are allowed to disappear only by pair annihilation.  If all the monopoles and antimonopoles
%that are created at the gap closing at $m=m_0$ are eventually pairwise annihilated at $m=m_1(>m_0)$, 
%the system again becomes an insulator in the bulk. 
%Let $m=m_1(>m_0)$ be the value of $m$ where the 
%system returns to an insulator again. 
%Such a gap-closing event
%is closely related to the formulae for the 
%$Z_2$ topological numbers, which are
% very different between inversion symmetric and inversion asymmetric systems. 

We apply this theory to BiTeI, which lacks inversion symmetry \cite{Ishizaka}. 
The space group of BiTeI is No.156 ($P3m1$). 
BiTeI is a NI at ambient pressure, and has been proposed to 
become a STI at high pressure \cite{Bahramy,BJYang}. 
Subsequently, {\it ab initio} calculations in a previous work \cite{Liu-BiTeI} showed existence of a Weyl semimetal phase between 
the NI and the STI phases in a narrow window of pressure, 
which had been overlooked in some previous works
\cite{Bahramy,BJYang}. 
%It is contrary to Ref.~\cite{Murakami07b}, and it is a critical issue how the 
%phase transition proposed in \cite{Murakami07b} is modified in the presence of
%additional spatial symmetries. 
When the pressure is increased, the gap first closes at six S points on the A-H lines, 
and six pairs of Weyl nodes are created.
The six monopoles and six antimonopoles move in opposite directions; subsequently, they annihilate each other at generic points on three mirror planes, leading to the STI phase \cite{Liu-BiTeI}. 
Combination of all the trajectories of the Weyl nodes yields a loop around the TRIM point (A point);
it means that a band inversion occurs at the A point between the low-pressure NI phase  and the high-pressure STI phase, and subsequent change of the $\mathbb{Z}_2$ topological number results
\cite{Murakami07b,MurakamiKuga}.
From our table, the space group No.156 at six S points gives ${\bf 1sp}$, meaning that  immediately 
after the pair creations, they move 
perpendicular to the AH line, in agreement with the previous work \cite{Liu-BiTeI}.

Other examples are LaBi$_{1-x}$Sb$_x$Te$_3$ and LuBi$_{1-x}$Sb$_x$Te$_3$, which 
has been proposed to undergo NI-WSM -TI phase transition
by changing $x$ \cite{Liu-BiTeI}. 
%The parent compounds  LaBiTe$_3$ and  LuBiTe$_3$ have a lattice structure similar
%to Bi$_2$Te$_3$ but the Bi layers are alternately replaced by La or Lu, thereby breaking
%the inversion symmetry. 
The space group is No.160 ($R3m$), lacking inversion symmetry. 
By increasing $x$ the band gap closes at
six generic points, corresponding to creation of Weyl nodes \cite{Liu-BiTeI}. 
The resulting 12 Weyl nodes move in ${\bf k}$ space, until they are annihilated at the 
six $B(\equiv \Sigma)$ points. 
In our table, the $B$ point corresponds to ${\bf 1sp}$, i.e. the pair annihilation occurs
between two Weyl nodes, which are moving perpendicular to the normal direction of the
mirror planes. It agrees with the previous work~\onlinecite{Liu-BiTeI}. 

{\bf Summary}--
To summarize, we investigate evolution of the band structure after 
parametric closing of the band gap in inversion-asymmetric crystals.
We found that there occur only two possibilities. One is the Weyl semimetal phase, and  closing of the gap 
corresponds to monopole-antimonopole 
pair creations of Weyl nodes. Distribution of the Weyl nodes in ${\bf k}$ space after the closing of the gap 
is uniquely determined by symmetry.
The other possibility is a nodal-line semimetal, with a gap closed along loops on mirror planes. 
From these results, we show that in any topological phase transitions with different $\mathbb{Z}_2$ topological numbers (without inversion symmetry), there should be a Weyl semimetal phase between 
the two bulk-insulating phases. 
These results give us 
a systematic 
way to search materials for topological semimetals.

{\small 
\noindent
{\bf Methods}\\
{\bf Details of the first-principles calculation}--
Electronic structure is obtained from the local density approximation (LDA) of the relativistic density functional theory (DFT). 
Calculation of the electronic structure is carried out using OpenMX code (http://www.openmx-square.org/)
 based on localized basis functions and norm-conserving pseudopotentials. 
We employ $12\times 12\times 12$ $\bm{k}$-point sampling for HfS, HfSe, and MgPt, 
and $6\times 6\times 12$ $\bm{k}$-point sampling for LuSI.
Lattice optimization for HfS under pressure is based on the LDA,
and is carried out using QMAS (Quantum MAterials Simulator) code (http://qmas.jp/) based on the projector augmented-wave method.
We employ $12\times 12\times 12$ $\bm{k}$-point sampling and 40 Ry as a plane-wave energy cutoff in the lattice optimization.

\noindent
{\bf Details of the effective-model calculation}--\\
It is convenient in the following analysis to expand the Hamiltonian as
$H({\bf q},m)=\sum_{i=1}^3 a_i({\bf q}, m)\sigma_i$, where $\sigma_i$ ($i=x,y,z$) are 
Pauli matrices, ${\bf q}={\bf k}-{\bf k}_0$. and $\sigma_z=+1$ and $-1$ correspond to
the conduction and the valence band, respectively. Here we omitted the term proportional to an identity matrix because it
does not affect the gap-closing event.  Since the gap closes at $m=m_0$, ${\bf q}=0$, 
we have $a_i({\bf q}=0, m=m_0)=0$, $i=1,2,3$. We assume that when $m<m_0$ the 
gap is open. In each case presented below, we examine whether the gap can close or not 
by counting a codimension $d_c$, 
i.e. the number of parameters to be tuned to close the gap. 
If the codimension is equal or lower than the number of tunable variables,
the gap can close there; otherwise the gap does not close.  

{\bf (i) no symmetry}: When there is no special symmetry 
at the ${\bf k}$ point, the Hamiltonian 
is not restricted by symmetry. 
The gap of the $2\times2$ Hamiltonian closes when $a_x=a_y=a_z=0$; this condition
determines a curve in the four-dimensional space $({\bf q},m)$. 
This curve goes through the point $({\bf q}=0, m=m_0)$, but it
does not exist in the $m<m_0$ region by assumption, and therefore the 
curve has a minimum value of $m$ at $m=m_0$ (Fig.~\ref{fig:mq} {\bf a}).
Therefore, as the value of $m$ increases, the gap-closing point 
appears at ${\bf q}=0$ for $m=m_0$ and then it 
splits into two points (Weyl nodes) when $m$ is increased further. It is 
a pair creation of a monopole and an antimonopole.

%aaaaaaaaaaaaaaaaaaaaaaaaaaaa

{\bf (ii) $C_{2}$ symmetry}: Suppose ${\bf k}$ is invariant only by a twofold ($C_2$) rotation, 
taken to be around the $z$ axis. 
For (ii-1): $R_{\rm c}=R_{\rm v}$,
the $2\times2$ Hamiltonian satisfies
%\begin{equation}
$H(q_x,q_y,q_z,m)=H(-q_x,-q_y,q_z,m)$,
%\end{equation}
where ${\bf q}\equiv {\bf k}-{\bf k}_0$. In particular, at ${\bf k}={\bf k}_0$, 
the above equation becomes trivial, imposing no constraint on the Hamiltonian.
Therefore, the gap cannot close by changing a single parameter $m$, because 
three 
conditions  $a_i({\bf q}=0,m)=0$, ($i=x,y,z$) cannot be simultaneously satisfied in general. 

On the other hand, for (ii-2): $R_{\rm c}\neq R_{\rm v}$, the opposite signs of the 
$C_2$ eigenvalues lead to an equation
%\begin{equation}
$H(q_x,q_y,q_z,m)=\sigma_z H(-q_x,-q_y,q_z,m)\sigma_z .$
%\end{equation}
In particular, for $q_x=q_y=0$, we obtain $H(0,0,q_z,m)=a_z(q_z,m)\sigma_z$, whose
gap closes when $a_z(q_z,m)=0$. From our assumption, 
it is satisfied when $q_z=0,m=m_0$ but 
is not for $m<m_0$. Hence, the value of $m$ satisfying $a_z(q_z,m)=0$ as a function of $q_z$ should have a minimum 
at $q_z=0$ (Fig.~\ref{fig:mq} {\bf b}).
It leads to bifurcation into
two solutions on the $q_z$ axis, as
$m$ increases across $m_0$, and it describes a
monopole-antimonopole pair creation of Weyl nodes. 
Therefore two Weyl nodes are created at $m=m_0$, and they move along the 
$C_2$ axis. 
We call this case as ``{\bf 1a}''. 
\begin{figure}[htp]
\includegraphics[width=7cm]{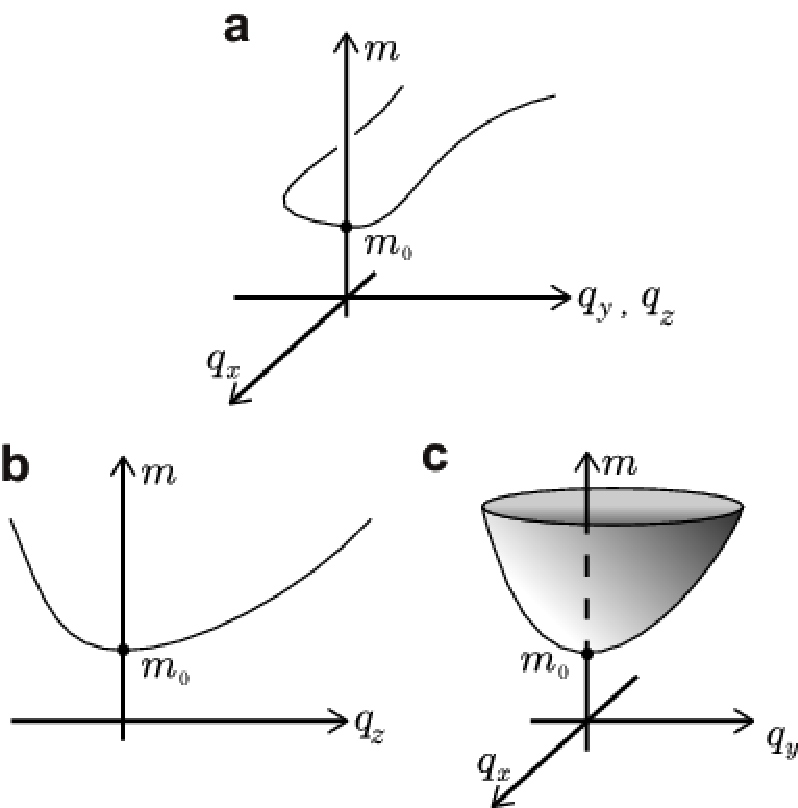}
\caption{\label{fig:mq} Values of $m$ for closing of the gap as a function of ${\bf q}$. 
{\bf a} 
$m=m(q_x, q_y, q_z)$ as a solution of $a_i(q_x,q_y,q_z,m)=0$ ($i=x,y,z$) in (i). {\bf b}, $m=m(q_z)$ that satisfies $a_z(0,0,q_z,m)=0$ in (ii-2). {\bf c}: 
$m=m(q_x,q_y)$ as a solution of $a_z(q_x,q_y,0,m)=0$ in (iv-2).}
\end{figure}

(iii) {\bf $C_2$ and $\Theta C_2$ symmetries}: 
Suppose the wavevector ${\bf k}$ is invariant 
under $C_{2z}$ and $\Theta C_{2y}$. 
%We set the $C_2$ axis to be the $z$ axis, and the
%$C'_2$ axis to be the $y$ axis. 
There are two eigenvalues of $C_{2z}$ with opposite signs, yielding two 
irreps.
For (iii-1) $R_{\rm c}=R_{\rm v}$, the Hamiltonian satisfies
%\begin{align}
$H(q_x,q_y,q_z,m)=H(-q_x,-q_y,q_z,m)=H^*(q_x,-q_y,q_z,m)$,
%\end{align}
where $*$ represents complex conjugation. These equations lead to 
$a_x=f_x(q^2_x,q_y^2,q_z,m)$, 
$a_y=q_xq_yf_y(q_x^2,q_y^2,q_z,m)$, 
$a_z=f_z(q_x^2,q_y^2,q_z,m)$. As $m$ is increased, it describes creations of two monopoles
and two antimonopoles at ${\bf q}=0$, $m=m_0$. For $m>m_0$, 
two monopoles are at $(\pm \tilde{q}_x,0,\tilde{q}_z)$ on 
the $xz$ plane and two antimonopoles are at $(0,\pm \tilde{q}_y,\tilde{q}'_z)$, where $\tilde{q}_x$, $\tilde{q}_y$, $\tilde{q}_z$ and 
$\tilde{q}'_z$ depend on $m$ (or the positions of 
the monopoles and antimonopoles might be exchanged). 
%One can see that this is the minimal configuration of monopoles and antimonopoles
%which satisfies the symmetry requirement; to see this we note that the spatial rotations and
%time reversal does not change the monopole charge, whereas the mirror operation flips the
%sign of the monopole charge. 
Notably, in contrast to (ii-1) where the level repulsion prohibits 
closing of the gap, 
in the present case (iii-1), the additional 
$\Theta C_2$ symmetry gives rise to additional constraints
and suppresses level repulsion.
%One can close the gap, but because of the level repulsion
% the gap immediately opens
%on the $C_2$ axis for $m>m_0$, and the Weyl nodes go
%away from the $C_2$ and $C'_2$ axes. 
In our list this behavior in (iii-1) is represented as
${\bf 2a}$. 

For (iii-2) $R_{\rm c}\neq R_{\rm v}$, the Hamiltonian satisfies 
%\begin{align}
$H(q_x,q_y,q_z,m)=\sigma_z H(-q_x,-q_y,q_z,m)\sigma_z=H^*(q_x,-q_y,q_z,m)$.
%\end{align}
The similar discussion as in (iii-1) leads to the result
$a_x=q_xf_x(q^2_x,q_y^2,q_z,m)$, 
$a_y=q_yf_y(q_x^2,q_y^2,q_z,m)$, 
$a_z=f_z(q_x^2,q_y^2,q_z,m)$. Therefore, for $q_x=q_y=0$, $a_x$ and $a_y$ identically vanish, 
and the remaining condition $a_z(0,0,q_z,m)=0$ describes a
pair creation of Weyl nodes at ${\bf q}=0$ and $m=m_0$ as in (ii-2). It is written as ${\bf 1a}$. 

{\bf (iv) mirror symmetry ${\cal M}$}:  let the $z$ axis denote the  
 direction normal to the mirror plane. 
Because a square of the mirror operation ${\cal M}$ is equal to 
$-1$ for mirror symmetry or an identity operation times a ${\bf k}$-dependent
factor for glide symmetry, there are two representations with opposite signs of mirror eigenvalues. 
For (iv-1): $R_{\rm c}=R_{\rm v}$, the Hamiltonian 
satisfies $H(q_x,q_y,q_z,m)=H(q_x,q_y,-q_z,m)$. 
This equation is automatically satisfied on the mirror plane $q_z=0$. Thus
closing of the gap on the mirror plane imposes three conditions $a_i(q_x,q_y,0,m)=0$, $i=x,y,z$
for three variables $q_x$, $q_y$, and $m$. This set of equations can have solutions 
on the mirror plane, and this closing of the gap accompanies a pair creation of Weyl nodes, which 
will then move symmetrically with respect to the mirror plane.
We call this pattern {\bf 1sa}.

On the other hand, for (iv-2): $R_{\rm c}\neq R_{\rm v}$, we obtain
%\begin{equation}
$H(q_x,q_y,q_z,m)=\sigma_z H(q_x,q_y,-q_z,m)\sigma_z$.
%\end{equation}
%By expanding $H$ as $H=\sum_{i}a_i({\bf k},m)\sigma_i$, the above equation means that
%$a_x$ and $a_y$ are odd functions of $q_z$. 
Hence, within the $q_z=0$ plane, the Hamiltonian is expressed as 
$H(q_x,q_z,0,m)=a_z(q_x,q_y,0,m)\sigma_z$. 
Closing of the gap imposes only one condition $a_z=0$ (i.e. $d_c=1$) for three 
variables ($q_x$, $q_y$, $m$); therefore, the gap can close by changing $m$. 
By assumption, 
the gap-closing condition, $a_z(q_x,q_y,0,m)=0$, does not have solutions for
$m<m_0$ and have a solution at $q_x=q_y=0, m=m_0$. 
Therefore, the value of $m$
as a function of $q_x$ and $q_y$ has a minimum at $q_x=q_y=0$ as shown in Fig.~\ref{fig:mq}{\bf c}.
When $m$ is larger than $m_0$, 
the gap closes along a loop (nodal line) within the $q_z=0$ plane.
We call this case ${\bf 1\bm{\ell}}$.

}

\noindent
{\bf Acknowledgement}\\
We thank Shoji Ishibashi for providing us with the ab initio code (QMAS) and pseudopotentials.
This work is partially supported by JSPS KAKENHI Grant Numbers JP26287062, JP26103006,
JP16J08552, and by MEXT Elements Strategy Initiative to Form Core Research Center (TIES). 

\vspace{2mm}

\noindent
{\bf Additional Information}\\
The authors declare no competing financial interests.

\vspace{2mm}

\noindent
{\bf Author contribution}\\
All authors contributed to the main contents of this work.
S.M. conceived and supervised the project. S.M. also did analytical calculations for
classifications in terms of the space groups with the help of R.O.  
M.H. performed  
the {\it ab initio} calculation with contributions from T.M.  
R.O. also constructed the theory on real materials such as BiTeI with the help of S.M.
M.H., R.O. and S.M. drafted the manuscript.
T.M. gave critical revisions of the manuscript. 

\clearpage

\noindent
{\Large Supplemental Materials}

\renewcommand{\thetable}{S\arabic{table}}
\renewcommand{\thefigure}{S\arabic{figure}}

\setcounter{figure}{0}

\section{More examples for gap closing with various symmetries}

In the main text we gave four examples for closing of the gap, classified by the 
${\bf k}$-group of the ${\bf k}$ point considered: (i) no symmetry, (ii) twofold rotation 
$C_2$, (iii) $C_2$ and $\Theta C'_2$,
and (iv) mirror reflection ${\cal M}$. Here we give more examples, in order to 
demonstrate evolution of the band structure as 
the value of the parameter $m$ changes.
As explained in the Methods section in the main text, 
we consider the Hamiltonian of the form
$H({\bf q},m)=\sum_{i=1}^3 a_i({\bf q}, m)\sigma_i$, where $\sigma_i$ ($i=x,y,z$) are 
Pauli matrices, ${\bf q}={\bf k}-{\bf k}_0$. We assume $a_i({\bf q}=0, m=m_0)=0$, $i=1,2,3$.
We also assume that when $m<m_0$ the gap is open. 

\begin{figure}[htp]
\includegraphics[width=7cm]{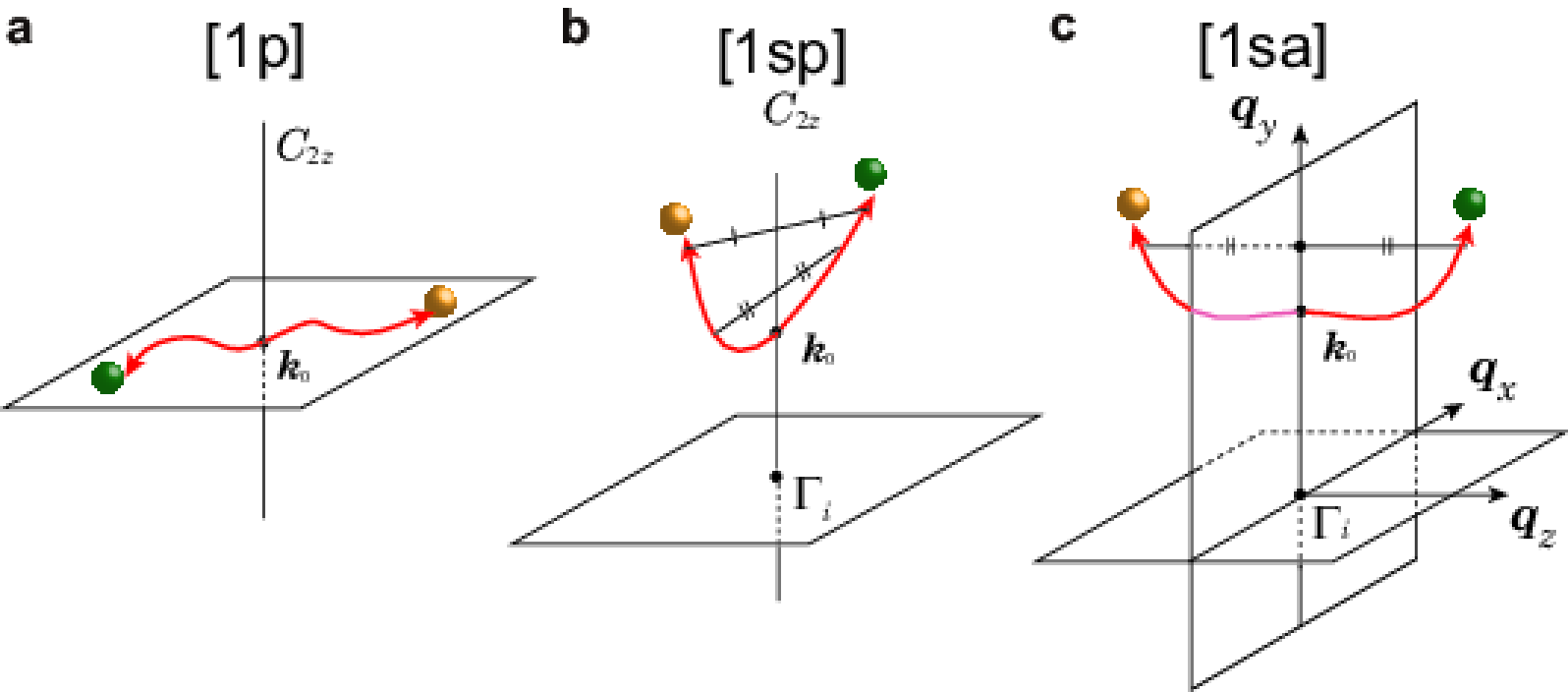}
\caption{\label{fig:supp1} Trajectories of Weyl nodes after a pair creation at ${\bf k}_0$ for
{\bf a}: (v), 
{\bf b}: (vi), and {\bf c}: (vii-1). Yellow and green spheres denote
monopoles and antimonopoles in ${\bf k}$ space, respectively, and they are both Weyl nodes. }
\end{figure}

(v) {\bf $\Theta C_2$ symmetry}: 
We consider a ${\bf k}$ point which is invariant only under an operation
$\Theta C_2$, where $\Theta$ is the time-reversal operation. Let $z$ denote a coordinate along the 
$C_2$ axis. The Hamiltonian then satisfies
$H(q_x,q_y,q_z,m)=H^*(q_x,q_y,-q_z,m)$. It leads to $a_y=q_z f_y(q_x,q_y,q_z^2,m)$, 
$a_x=f_x(q_x,q_y,q_z^2,m)$, and $a_z=f_z(q_x,q_y,q_z^2,m)$, where $f_i$ are analytic functions. From these conditions, we can show that 
 the solution at ${\bf q}=0$,
$m=m_0$ bifurcates into two Weyl nodes, a monopole and an antimonopole on 
the $q_z=0$ plane, and their trajectories are given by $q_z=0$, $f_x(q_x,q_y,0,m)=0=f_z(q_x,q_y,0,m)$. This case is denoted as ${\bf 1p}$, where ${\bf p}$ stands for ``planar'', 
meaning that the trajectories of the monopole and the antimonopole are 
restricted to be within a high-symmetry plane (Fig.~\ref{fig:supp1}{\bf a}). 

(vi) {\bf $\Theta {\cal M}$ symmetry}: 
Suppose the wavevector ${\bf k}$ is invariant under $\Theta{\cal M}$, where ${\cal M}$ is a mirror operation. We take the $z$ axis 
to be perpendicular to the mirror plane. We then obtain $H(q_x,q_y,q_z,m)=H^*(-q_x,-q_y,q_z,m)$, leading 
to $a_y=q_xf_x(q_x^2,q_y^2,q_z,m)+q_yf_y(q_x^2,q_y^2,q_z,m)$, while $a_x$ and $a_z$ are
analytic functions of $q_x^2, q_y^2,q_z$ and $m$. By retaining the lowest order in the arguments in these functions, 
one can see that it describes a monopole-antimonopole pair creation and the positions of the monopole and the antimonopole are 
expressed as $
(\tilde{q}_x(m), \tilde{q}_y(m), \tilde{q}_z(m))$ and 
$(-\tilde{q}_x(m), -\tilde{q}_y(m), \tilde{q}_z(m))$,  respectively. They are
symmetric with respect to the 
$q_x=q_y=0$ axis. We call this case as ${\bf 1sp}$, where ${\bf s}$ (symmetric) means
that the positions of the monopole and the antimonopole are related with each 
other by symmetry operations  (Fig.~\ref{fig:supp1}{\bf b}).  

(vii) {\bf ${\cal M}$ and $\Theta {\cal M}'$ symmetries, with mirror planes of ${\cal M}$ and ${\cal M}'$ perpendicular to
each other}:
Let us call the mirror planes for ${\cal M}$ and $\Theta {\cal M}'$ as $xy$ and $xz$ planes, 
respectively. There are two representations corresponding to different signs of the 
eigenvalues of ${\cal M}$.  Evolution of band structure after closing of the gap is different between (vii-1) 
the two bands with the same irreducible representations ($R_{\rm c}=R_{\rm v}$), and (vii-2) those with 
different irreducible representations  ($R_{\rm c}\neq R_{\rm v}$) . For (vii-1), we have
\begin{align*}
&H(q_x,q_y,q_z,m)=H(q_x,q_y,-q_z,m),\\
&H(q_x,q_y,q_z,m)=H^*(-q_x,q_y,-q_z,m),\end{align*}
leading to 
$a_x=f_x(q^2_x,q_y,q_z^2,m)$, 
$a_y=q_xf_y(q_x^2,q_y,q_z^2,m)$, 
$a_z=f_z(q_x^2,q_y,q_z^2,m)$.
Therefore, for $q_x=0$, the gap closes when $f_x(0,q_y,q_z^2,m)=0=f_z(0,q_y,q_z^2,m)$ is satisfied. 
These two equations determine a curve in the $(q_y, q_z,m)$ space, and 
this curve is symmetric with respect to $q_z\rightarrow -q_z$. 
This curve passes through the point $q_y=0$, $q_z=0$, $m=m_0$, but does not 
go  into the $m<m_0$ region. 
Therefore, as we increase $m$ across $m_0$, the 
solution at $q_y=0=q_z, m=m_0$ bifurcates into a pair of Weyl nodes at ${\bf q}=(0, \tilde{q}_y(m),
\pm\tilde{q}_z(m))$. It is denoted as {\bf 1sa}  (Fig.~\ref{fig:supp1}{\bf c}). For the case 
 (vii-2) $R_{\rm c}\neq R_{\rm v}$, because the mirror symmetry persists for all the ${\bf k}$
points within the mirror plane ($xy$ plane), it gives rise to a nodal-line semimetal after closing of the gap, denoted as 1$\bm{\ell}$.

\section{Classification tables for emergent topological semimetals after closing of the band gap of inversion-asymmetric insulators}
\begin{figure*}[htp]
\includegraphics[width=15cm]{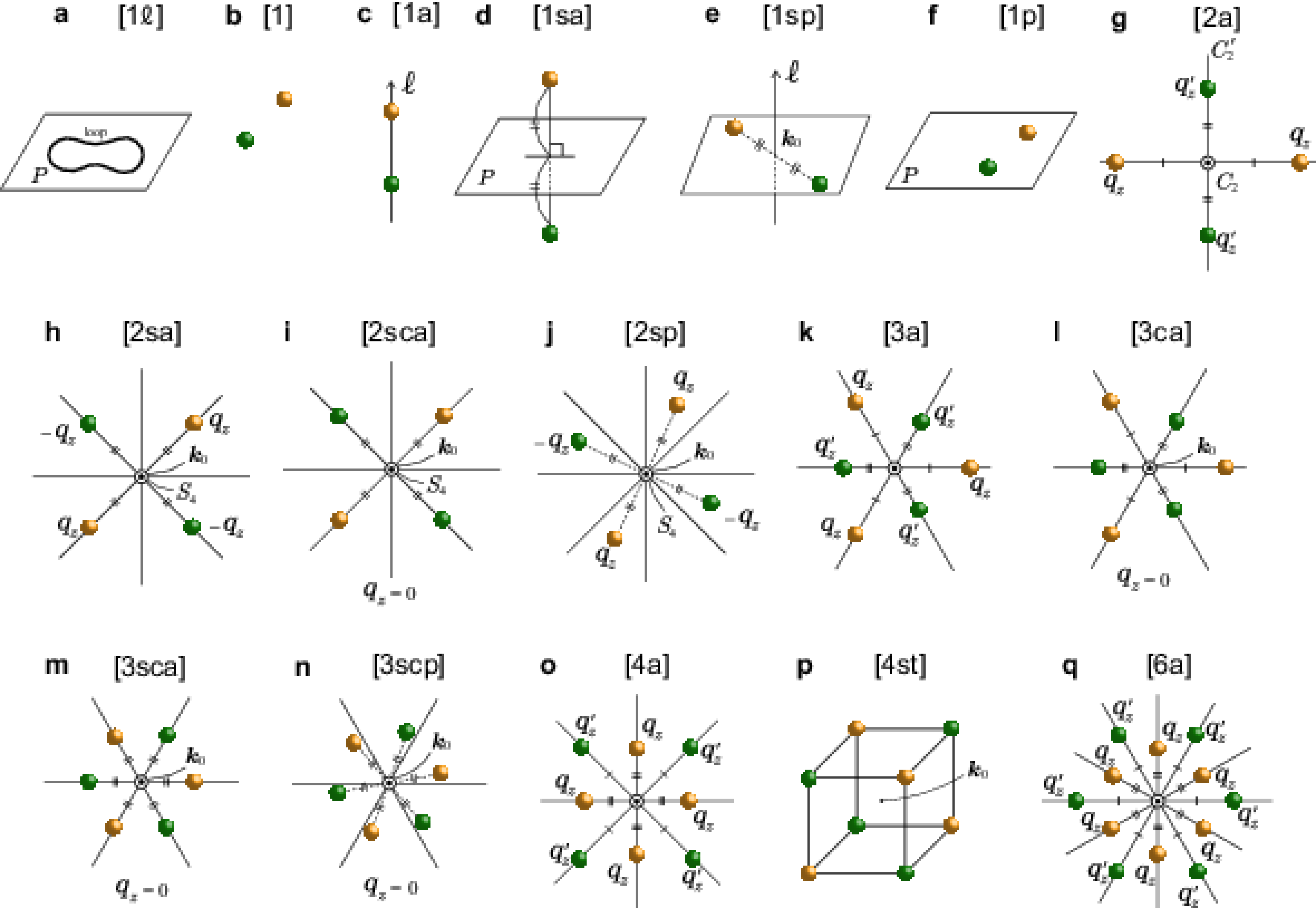}
\caption{\label{fig:S3} All the patterns of gap-closing points after parametric gap closing. 
{\bf a}, Line nodes for a nodal-line semimetal. It shows a single line node, whereas the number of line nodes can range among 1, 2, 3, 4 and 6. {\bf b}-{\bf q}, All the pattern of Weyl nodes after parametric gap closing at ${\bf k}_0$. Yellow and green spheres denote
monopoles and antimonopoles in ${\bf k}$ space, respectively, and they are both Weyl nodes. 
Solid lines 
denote high-symmetry directions. 
The $z$ axis is parallel to a high-symmetry axis which is
taken to be perpendicular to the page, and 
${\bf q}\equiv {\bf k}-{\bf k}_0$ is the 
momentum measured from ${\bf k}_0$. In {\bf i}, {\bf l}, {\bf m} and {\bf n}, all the 
Weyl nodes are coplanar with ${\bf k}_0$, while in {\bf g}, {\bf h}, {\bf j}, {\bf k}, {\bf o},  and {\bf q}, the Weyl nodes are mutually displaced along the $z$ direction.
}
\end{figure*}

In the main text, we have shown that patterns for emergence of Weyl nodes and nodal lines after closing of the gap
are uniquely determined by symmetry, i.e. by the space group and by the value of ${\bf k}_0$ 
where the gap closes. 
Here all the patterns for Weyl nodes and nodal lines for all the 138 space groups are shown in  Fig.~\ref{fig:S3}.
While this figure is the same as Fig.~4 in the main text, and it is shown here again for 
readers' convenience.
Notations used in this figure are summarized as follows.
\begin{itemize}
\item
The numbers in the symbols in e.g. {\bf 1s} and {\bf 1sp} represent the number of pair of Weyl nodes, except for ${\bf 1\bm{\ell}, 2\bm{\ell},3\bm{\ell}, 4\bm{\ell}}$, 
where the number represents a number of nodal lines. 
\item The symbol ${\bf \bm{\ell}}$ means that there 
is a nodal line where the gap is closed. Such a nodal line always appears on mirror planes, 
and only when  the valence and the conduction bands have different 
mirror eigenvalues. 
\item
The symbol ${\bf a}$ (axial) 
represents that relative directions between the Weyl nodes are fixed to be along certain high-symmetry lines.
\item
 The symbol ${\bf p}$ (planar) represents that relative directions between the Weyl nodes are not confined to be
a high-symmetry axis, but confined within a high-symmetry plane. 
\item
The symbol ${\bf c}$ (coplanar) means that all the monopoles and antimonopoles 
lie on a same high-symmetry plane. This symbol ${\bf c}$ is used only when there are more than one 
monopole-antimonopole pairs. 
\item
The symbol ${\bf t}$ (tetrahedron) appears only for 
a few cases with tetrahedral or cubic symmetries. It means that 
the 4 monopoles and 4 antimonopoles form 8 vertices of a cube whose center is a high 
symmetry point, and 4 monopoles form a tetrahedron. 
\item
The symbol ${\bf s}$ (symmetric) means that all the monopoles and antimonopoles are related to each other 
by symmetry operations. In such cases
they are energetically degenerate, and it is possible to locate all the Weyl nodes 
on the Fermi energy. Otherwise, these monopoles and antimonopoles may not necessarily be at
the same energy. 
\end{itemize}

Apart from these patterns in Fig.~\ref{fig:S3}, their combinations can appear, such as
${\bf 1sa1\bm{\ell}}$ (e.g. at K and H points in No.174 and No.187), which 
means that one pair of Weyl nodes and one nodal-line are simultaneously 
created at closing of the gap. In addition, when there are more than one mirror planes 
at ${\bf k}_0$, more than one nodal lines can be created 
(e.g. ${\bf 2\bm{\ell}}$, ${\bf 3\bm{\ell}}$, ${\bf 4\bm{\ell}}$), and they are on different
mirror planes, intersecting each other at ${\bf k}_0$.

In the following, we explain how these patterns emerge for various ${\bf k}$ points in 138 space groups
without inversion symmetry. Time-reversal symmetry is assumed throughout. 
All the cases are classified in terms of space groups and values of ${\bf k}_0$ where the 
gap closes. 
Then, various cases for each ${\bf k}$ point are classified in terms of 
irreducible representation of the conduction band and that of the valence band, 
denoted as $R_c$ and $R_v$, respectively. As explained in the main text, we have to study only the cases with ${\rm dim} R_c={\rm dim} R_v=1$ because otherwise the gap cannot close ${\bf k}_0$. In particular, one can exclude the TRIM as ${\bf k}_0$, since there is always Kramers degeneracy at the TRIM, and the gap cannot close.  
In addition, if a high-symmetry point shares the same $k$-group with a
high-symmetry line which includes this point, the gap does not close at the
high-symmetry point, because it requires fine-tuning of the Hamiltonian.

In the following tables, we  show the high-symmetry points and lines, where the gap can close 
by changing a single parameter $m$. 
It should be noted that in addition to the patterns listed in the following tables, the following 
two patterns exist, which are common to all the space groups.
\begin{itemize}
\item generic ${\bf k}$ without any symmetry: ${\bf 1}$ (the case (i) in the main text)
\item generic ${\bf k}$ on a mirror plane: 
{\bf 1sa} for $R_{\rm c}= R_{\rm v}$ (the case (iv-1)), and 
${\bf 1\ell}$, 
for $R_{\rm c}\neq R_{\rm v}$ (the case (iv-2)). In the former case, 
the conduction and valence bands
have the same signs of mirror eigenvalues, while in the latter, they have the opposite signs.
\end{itemize}

In the following tables \ref{tab:S1}-\ref{tab:S5}, we show all the patterns for the combinations 
of $R_c$ and $R_v$, where the gap can close by changing the parameter $m$. Cases where the gap cannot close at ${\bf k}_0$ 
are excluded from the 
tables.
We show all the 138 space groups without inversion symmetry. In the tables 
``point'' and ``line'' refer to the high-symmetry points and the high-symmetry lines, respectively. As noted above, $R_c$ and $R_v$ should be both one-dimensional. 
The names of the irreducible representations (irreps) follows the notation in Ref. \onlinecite{BradleyS}, such as
$R_1$ and $R_2$, and so on.  
For a concise expression in the tables, we use the symbols
``ii'' and ``ij''; 
``ii'' means the case with $R_{\rm c}=R_{\rm v}$ and
``ij'' means the case with $R_{\rm c}\neq R_{\rm v}$, respectively. 
When there is only one representation allowed for the ${\bf k}$ point,
such symbols are not needed and thus are not shown.
For more complicated cases, for example, the entry $(3,4)$ in the tables 
means 
 $(R_{\rm c},R_{\rm v})=(R_3,R_4),(R_4,R_3)$.
Furthermore, some lengthy entities are abbreviated in the following way:\\
$\langle 1 \rangle$=($R_2$,$R_3$)($R_4$,$R_5$); {\bf 2$\bm{\ell}$}, ($R_2$,$R_4$)($R_3$,$R_5$)($R_2$,$R_5$)($R_3$,$R_4$); {\bf 1$\bm{\ell}$}.\\
$\langle 2 \rangle$ = ($R_5$,$R_6$)($R_6$,$R_7$)($R_7$,$R_8$)($R_8$,$R_5$); {\bf 1sa}, ($R_5$,$R_7$)($R_6$,$R_8$); {\bf 2sp}.\\
$\langle 3 \rangle$=($R_2$,$R_5$)($R_3$,$R_4$); {\bf 2$\bm{\ell}$},($R_2$,$R_3$)($R_4$,$R_5$)($R_2$,$R_4$)($R_3$,$R_5$); {\bf 1$\bm{\ell}$}.\\
$\langle 4 \rangle$ =($R_5$,$R_6$)($R_6$,$R_7$)($R_7$,$R_8$)($R_8$,$R_5$); {\bf 1sa} ($R_5$,$R_7$)($R_6$,$R_8$); {\bf 2sa}.\\
$\langle 5 \rangle$ = ($R_3$,$R_4$)($R_5$,$R_6$); {\bf 2$\bm{\ell}$} ($R_3$,$R_5$)($R_4$,$R_6$); {\bf 2sca} ($R_3$,$R_6$)($R_4$,$R_5$); {\bf 1$\bm{\ell}$}\\
$\langle 6 \rangle$ = 
($R_{11}$,$R_8$)($R_{10}$,$R_9$)($R_7$,$R_{12}$); {\bf 1$\bm{\ell}$}, ($R_{11}$,$R_7$)($R_{10}$,$R_8$)($R_7$,$R_9$) ($R_8$,$R_{12}$)($R_9$,$R_{11}$)($R_{12}$,$R_{10}$); {\bf 1sa}, \\
($R_{11}$,$R_{10}$)($R_{10}$,$R_7$)($R_7$,$R_8$)($R_8$,$R_9$)($R_9$,$R_{12}$)($R_{12}$,$R_{11}$); {\bf 1sa1$\bm{\ell}$}.\\
$\langle 7 \rangle$=($R_4$,$R_5$)($R_5$,$R_6$)($R_6$,$R_4$); {\bf 4st}.\\
$\langle 8 \rangle$=($R_4$,$R_7$)($R_5$,$R_6$); {\bf 4$\bm{\ell}$},($R_4$,$R_5$)($R_6$,$R_7$); 
{\bf 3$\bm{\ell}$},($R_4$,$R_6$)($R_5$,$R_7$); {\bf 1$\bm{\ell}$}.\\

\begin{table}[h]
\centering
{\footnotesize
\begin{tabular}{cc|c|c} 
\multicolumn{2}{c|}{space group}& point & line\\
\hline
{\bf 1} & {\it P}1 &  & \\
{\bf 3} & {\it P}2 &  & $\Lambda$VWU:ij;{\bf 1a}\\
{\bf 4} & {\it P}$2_{1}$ &  & $\Lambda$VWU:ij;{\bf 1a}\\
{\bf 5} & {\it C}2 &  & $\Lambda$U:ij;{\bf 1a}\\
{\bf 6} & {\it Pm} &  & $\Lambda$VWU:{\bf 1sp}\\
{\bf 7} & {\it Pc} &  & $\Lambda$W:{\bf 1sp}\\
{\bf 8} & {\it Cm} &  & $\Lambda$U:{\bf 1sp}\\
{\bf 9} & {\it Cc} &  & $\Lambda$:{\bf 1sp}\\
\end{tabular}}
\caption{\label{tab:S1}Patterns of gap-closing points after parametric gap closing for  triclinic and monoclinic space groups.}
\end{table}

\section{Properties of band structure of HfS}
Here we briefly explain the band structure of HfS and the reason 
why it is appropriate as a candidate for a nodal-line semimetal.
The crystal structure of HfS is the same as that of tungsten carbide (WC).
Its space group is No. 187 ($P$-$6m2$)~\cite{FranzenS}, and 
it has several mirror planes,
which are necessary for the Weyl nodal lines by mirror symmetry as we discussed.
Bands of the Hf $5d$ orbitals are classified into three groups, $a_1'$, $e'$, and $e''$, by the crystal field from a trigonal prism of 
S$^{2+}$ atoms.
A narrow gap is formed between the occupied $a_1'$ band and the unoccupied $e'$ bands.
A layered triangle lattice formed by Hf atoms is favorable for 
displacing the point of the minimum band gap away from the $\Gamma$ point to the K point by interference effects.
When the spin-orbit coupling (SOC) is neglected, Dirac nodal-line exists around the K point on the $k_z =0$ plane.

When the SOC is considered, due to mismatch between threefold rotational symmetry of the trigonal prism and the right angle of 
electronic clouds of atomic $d$ orbitals, wavefunctions 
become necessarily complex, and 
a relativistic energy splitting would be partially enhanced. 
Spin splitting vanishes at the TRIM, while it is prominent at the non-TRIM points such as K- and H-points.
Thus the SOC lifts the degeneracy in the Dirac nodal-line, and the Weyl nodal-line appears instead near the K point on the mirror plane 
$k_z =0$.

\begin{table}[H]
\centering
{\footnotesize
\begin{tabular}{cc|c|c}
\multicolumn{2}{c|}{space group}& point & line\\
 \hline
{\bf 16} & {\it P}222 &  & $\Delta$DPB$\Sigma$CEA$\Lambda$HQG:ii;{\bf 2a},ij;{\bf 1a}\\
{\bf 17} & {\it P}$222_{1}$ &  & $\Delta$D$\Sigma$C$\Lambda$HQG:ii;{\bf 2a},ij;{\bf 1a}\\
{\bf 18} & {\it P}$2_{1}2_{1}2$ &  & $\Delta$B$\Sigma$A$\Lambda$:ii;{\bf 2a},ij;{\bf 1a}\\
{\bf 19} & {\it P}$2_{1}2_{1}2_{1}$ &  & $\Delta$$\Sigma$$\Lambda$:ii;{\bf 2a},ij;{\bf 1a}\\
{\bf 20} & {\it C}$222_{1}$ &  & $\Lambda$H$\Sigma$$\Delta$FC:ii;{\bf 2a},ij;{\bf 1a} D:ij;{\bf 1a}\\
{\bf 21} & {\it C}222 &  & $\Lambda$HA$\Sigma$$\Delta$BGFEC:ii;{\bf 2a},ij;{\bf 1a} D:ij;{\bf 1a}\\
{\bf 22} & {\it F}222 &  & $\Lambda$GHQ$\Sigma$CAU$\Delta$DBR:ii;{\bf 2a},ij;{\bf 1a}\\
{\bf 23} & {\it I}222 &  & $\Lambda$G$\Sigma$F$\Delta$U:ii;{\bf 2a},ij;{\bf 1a} PDQ:ij;{\bf 1a}\\
{\bf 24} & {\it I}$2_{1}2_{1}2_{1}$ & W:ij;{\bf 2a} & $\Lambda$G$\Sigma$F$\Delta$U:ii;{\bf 2a},ij;{\bf 1a} PDQ:ij;{\bf 1a}\\
{\bf 25} & {\it Pmm}2 &  & $\Delta$DPB$\Sigma$CEA:ii;{\bf 1sa},ij;{\bf 1$\bm{\ell}$}\\
{\bf 26} & {\it Pmc}$2_{1}$ &  & $\Delta$D$\Sigma$C:ii;{\bf 1sa},ij;{\bf 1$\bm{\ell}$}\\
{\bf 27} & {\it Pcc}2 &  & $\Delta$D$\Sigma$C:ii;{\bf 1sa},ij;{\bf 1$\bm{\ell}$}\\
{\bf 28} & {\it Pma}2 &  & $\Delta$DPB$\Sigma$A:ii;{\bf 1sa},ij;{\bf 1$\bm{\ell}$} HQ:$\langle 1 \rangle$\\
{\bf 29} & {\it Pca}$2_{1}$ &  & $\Delta$D$\Sigma$:ii;{\bf 1sa},ij;{\bf 1$\bm{\ell}$} HQ:$\langle 1 \rangle$\\
{\bf 30} & {\it Pnc}2 &  & $\Delta$D$\Sigma$E:ii;{\bf 1sa},ij;{\bf 1$\bm{\ell}$} HQ:$\langle 1 \rangle$\\
{\bf 31} & {\it Pmn}$2_{1}$ &  & $\Delta$D$\Sigma$:ii;{\bf 1sa},ij;{\bf 1$\bm{\ell}$} HQ:$\langle 1 \rangle$\\
{\bf 32} & {\it Pba}2 &  & $\Delta$B$\Sigma$A:ii;{\bf 1sa},ij;{\bf 1$\bm{\ell}$} HG:$\langle 1 \rangle$\\
{\bf 33} & {\it Pna}$2_{1}$ &  & $\Delta$$\Sigma$:ii;{\bf 1sa},ij;{\bf 1$\bm{\ell}$} HG:$\langle 1 \rangle$\\
{\bf 34} & {\it Pnn}2 &  & $\Delta$P$\Sigma$E:ii;{\bf 1sa},ij;{\bf 1$\bm{\ell}$} HG:$\langle 1 \rangle$\\
{\bf 35} & {\it Cmm}2 &  & D:ij;{\bf 1a} A$\Sigma$$\Delta$BGFEC:ii;{\bf 1sa},ij;{\bf 1$\bm{\ell}$}\\
{\bf 36} & {\it Cmc}$2_{1}$ &  & D:ij;{\bf 1a} $\Sigma$$\Delta$FC:ii;{\bf 1sa},ij;{\bf 1$\bm{\ell}$}\\
{\bf 37} & {\it Ccc}2 &  & D:ij;{\bf 1a} $\Sigma$$\Delta$FC:ii;{\bf 1sa},ij;{\bf 1$\bm{\ell}$}\\
{\bf 38} & {\it Amm}2 &  & $\Lambda$HA$\Sigma$EC:ii;{\bf 1sa},ij;{\bf 1$\bm{\ell}$} D:{\bf 1sp}\\
{\bf 39} & {\it Aem}2 &  & $\Lambda$HA$\Sigma$EC:ii;{\bf 1sa},ij;{\bf 1$\bm{\ell}$}\\
{\bf 40} & {\it Ama}2 &  & $\Lambda$H$\Sigma$C:ii;{\bf 1sa},ij;{\bf 1$\bm{\ell}$} D:{\bf 1sp} BG:$\langle 3 \rangle$\\
{\bf 41} & {\it Aea}2 &  & $\Lambda$H$\Sigma$C:ii;{\bf 1sa},ij;{\bf 1$\bm{\ell}$} BG:$\langle 3 \rangle$\\
{\bf 42} & {\it Fmm}2 &  & $\Sigma$CAU$\Delta$DBR:ii;{\bf 1sa},ij;{\bf 1$\bm{\ell}$}\\
{\bf 43} & {\it Fdd}2 &  & $\Sigma$U$\Delta$R:ii;{\bf 1sa},ij;{\bf 1$\bm{\ell}$} GH:$\langle 1 \rangle$\\
{\bf 44} & {\it Imm}2 & W:ij;{\bf 1sa} & P:ij;{\bf 1a} $\Sigma$F$\Delta$U:ii;{\bf 1sa},ij;{\bf 1$\bm{\ell}$} DQ:{\bf 1sp}\\
{\bf 45} & {\it Iba}2 &  & P:ij;{\bf 1a} $\Sigma$F$\Delta$U:ii;{\bf 1sa},ij;{\bf 1$\bm{\ell}$}\\
{\bf 46} & {\it Ima}2 &  & P:ij;{\bf 1a} $\Sigma$F$\Delta$U:ii;{\bf 1sa},ij;{\bf 1$\bm{\ell}$} Q:{\bf 1sp}\\
\end{tabular}
}
\caption{\label{tab:S2}Patterns of gap-closing points after parametric gap closing for  orthorhombic space groups.
We note that in No38, 39, 40 and 41, despite the space group symbols use the $A$ lattice 
for orthorhombic base-centered Bravais lattice, we use the $C$ lattice with the twofold rotation axis is along $y$, following the notations in Ref.~\onlinecite{BradleyS} (see p.83 and p135 
in Ref.~\onlinecite{BradleyS})}
\end{table}

\newpage

\begin{table}[H]
\centering
{\footnotesize
\begin{tabular}{cc|c|c} 
\multicolumn{2}{c|}{space group}& point & line\\
\hline
{\bf 75} & {\it P}4 &  & $\Delta$U$\Sigma$SYT:{\bf 1p} $\Lambda$VW:ij;{\bf 1a}\\
{\bf 76} & {\it P}$4_{1}$ &  & $\Delta$$\Sigma$Y:{\bf 1p} $\Lambda$VW:ij;{\bf 1a}\\
{\bf 77} & {\it P}$4_{2}$ &  & $\Delta$U$\Sigma$SYT:{\bf 1p} $\Lambda$VW:ij;{\bf 1a}\\
{\bf 78} & {\it P}$4_{3}$ &  & $\Delta$$\Sigma$Y:{\bf 1p} $\Lambda$VW:ij;{\bf 1a}\\
{\bf 79} & {\it I}4 &  & $\Lambda$VW:ij;{\bf 1a} $\Sigma$F$\Delta$UY:{\bf 1p}\\
{\bf 80} & {\it I}$4_{1}$ &  & $\Lambda$VW:ij;{\bf 1a} $\Sigma$F$\Delta$UY:{\bf 1p}\\
{\bf 81} & {\it P}$\overline{4}$ &  & $\Delta$U$\Sigma$SYT:{\bf 1p} W:ij;{\bf 1a}\\
{\bf 82} & {\it I}$\overline{4}$ & P:$\langle 2 \rangle$ & W:ij;{\bf 1a} $\Sigma$F$\Delta$UY:{\bf 1p}\\
{\bf 89} & {\it P}422 &  & $\Delta$U$\Sigma$SYTW:ii;{\bf 2a},ij;{\bf 1a} $\Lambda$V:ii;{\bf 4a},ij;{\bf 1a}\\
{\bf 90} & {\it P}$42_{1}$2 &  & $\Delta$U$\Sigma$S:ii;{\bf 2a},ij;{\bf 1a} $\Lambda$:ii;{\bf 4a},ij;{\bf 1a}\\
{\bf 91} & {\it P}$4_{1}$22 &  & $\Delta$$\Sigma$YW:ii;{\bf 2a},ij;{\bf 1a} $\Lambda$V:ii;{\bf 4a},ij;{\bf 1a}\\
{\bf 92} & {\it P}$4_{1}2_{1}$2 &  & $\Delta$$\Sigma$:ii;{\bf 2a},ij;{\bf 1a} $\Lambda$:ii;{\bf 4a},ij;{\bf 1a}\\
{\bf 93} & {\it P}$4_{2}$22 &  & $\Delta$U$\Sigma$SYTW:ii;{\bf 2a},ij;{\bf 1a} $\Lambda$V:ii;{\bf 4a},ij;{\bf 1a}\\
{\bf 94} & {\it P}$4_{2}2_{1}$2 &  & $\Delta$U$\Sigma$S:ii;{\bf 2a},ij;{\bf 1a} $\Lambda$:ii;{\bf 4a},ij;{\bf 1a}\\
{\bf 95} & {\it P}$4_{3}$22 &  & $\Delta$$\Sigma$YW:ii;{\bf 2a},ij;{\bf 1a} $\Lambda$V:ii;{\bf 4a},ij;{\bf 1a}\\
{\bf 96} & {\it P}$4_{3}2_{1}$2 &  & $\Delta$$\Sigma$:ii;{\bf 2a},ij;{\bf 1a} $\Lambda$:ii;{\bf 4a},ij;{\bf 1a}\\
{\bf 97} & {\it I}422 &  & $\Lambda$V:ii;{\bf 4a},ij;{\bf 1a} W$\Sigma$F$\Delta$UY:ii;{\bf 2a},ij;{\bf 1a} Q:ij;{\bf 1a}\\
{\bf 98} & {\it I}$4_{1}$22 & P:(2,3);{\bf 4a} & $\Lambda$V:ii;{\bf 4a},ij;{\bf 1a} W$\Sigma$F$\Delta$UY:ii;{\bf 2a},ij;{\bf 1a} Q:ij;{\bf 1a}\\
{\bf 99} & {\it P}4{\it mm} &  & $\Delta$U$\Sigma$SYT:ii;{\bf 1sa},ij;{\bf 1$\bm{\ell}$}\\
{\bf 100} & {\it P}4{\it bm} &  & $\Delta$U$\Sigma$S:ii;{\bf 1sa},ij;{\bf 1$\bm{\ell}$} W:$\langle 3 \rangle$\\
{\bf 101} & {\it P}$4_{2}${\it cm} &  & $\Delta$$\Sigma$SY:ii;{\bf 1sa},ij;{\bf 1$\bm{\ell}$}\\
{\bf 102} & {\it P}$4_{2}${\it nm} &  & $\Delta$$\Sigma$ST:ii;{\bf 1sa},ij;{\bf 1$\bm{\ell}$} W:$\langle 3 \rangle$\\
{\bf 103} & {\it P}4{\it cc} &  & $\Delta$$\Sigma$Y:ii;{\bf 1sa},ij;{\bf 1$\bm{\ell}$}\\
{\bf 104} & {\it P}4{\it nc} &  & $\Delta$$\Sigma$T:ii;{\bf 1sa},ij;{\bf 1$\bm{\ell}$} W:$\langle 3 \rangle$\\
{\bf 105} & {\it P}$4_{2}${\it mc} &  & $\Delta$U$\Sigma$YT:ii;{\bf 1sa},ij;{\bf 1$\bm{\ell}$}\\
{\bf 106} & {\it P}$4_{2}${\it bc} &  & $\Delta$U$\Sigma$:ii;{\bf 1sa},ij;{\bf 1$\bm{\ell}$} W:$\langle 3 \rangle$\\
{\bf 107} & {\it I}4{\it mm} &  & $\Sigma$F$\Delta$UY:ii;{\bf 1sa},ij;{\bf 1$\bm{\ell}$} Q:{\bf 1sp}\\
{\bf 108} & {\it I}4{\it cm} &  & $\Sigma$F$\Delta$UY:ii;{\bf 1sa},ij;{\bf 1$\bm{\ell}$}\\
{\bf 109} & {\it I}$4_{1}${\it md} & P:(13,14);{\bf 2$\bm{\ell}$} & W:$\langle 3 \rangle$ $\Sigma$F$\Delta$:ii;{\bf 1sa},ij;{\bf 1$\bm{\ell}$} Q:{\bf 1sp}\\
{\bf 110} & {\it I}$4_{1}${\it cd} &  & W:$\langle 3 \rangle$ $\Sigma$F$\Delta$:ii;{\bf 1sa},ij;{\bf 1$\bm{\ell}$}\\
{\bf 111} & {\it P}$\overline{4}$2{\it m} &  & $\Delta$UYTW:ii;{\bf 2a},ij;{\bf 1a} $\Sigma$S:ii;{\bf 1sa},ij;{\bf 1$\bm{\ell}$}\\
{\bf 112} & {\it P}$\overline{4}$2{\it c} &  & $\Delta$UYTW:ii;{\bf 2a},ij;{\bf 1a} $\Sigma$:ii;{\bf 1sa},ij;{\bf 1$\bm{\ell}$}\\
{\bf 113} & {\it P}$\overline{4}$$2_{1}${\it m} &  & $\Delta$U:ii;{\bf 2a},ij;{\bf 1a} $\Sigma$S:ii;{\bf 1sa},ij;{\bf 1$\bm{\ell}$}\\
{\bf 114} & {\it P}$\overline{4}$$2_{1}${\it c} &  & $\Delta$U:ii;{\bf 2a},ij;{\bf 1a} $\Sigma$:ii;{\bf 1sa},ij;{\bf 1$\bm{\ell}$}\\
{\bf 115} & {\it P}$\overline{4}${\it m}2 &  & $\Delta$UYT:ii;{\bf 1sa},ij;{\bf 1$\bm{\ell}$} $\Sigma$S:ii;{\bf 2a},ij;{\bf 1a}\\
{\bf 116} & {\it P}$\overline{4}${\it c}2 &  & $\Delta$Y:ii;{\bf 1sa},ij;{\bf 1$\bm{\ell}$} $\Sigma$S:ii;{\bf 2a},ij;{\bf 1a}\\
{\bf 117} & {\it P}$\overline{4}${\it b}2 &  & $\Delta$U:ii;{\bf 1sa},ij;{\bf 1$\bm{\ell}$} $\Sigma$S:ii;{\bf 2a},ij;{\bf 1a} W:$\langle 3 \rangle$\\
{\bf 118} & {\it P}$\overline{4}${\it n}2 &  & $\Delta$T:ii;{\bf 1sa},ij;{\bf 1$\bm{\ell}$} $\Sigma$S:ii;{\bf 2a},ij;{\bf 1a} W:$\langle 3 \rangle$\\
{\bf 119} & {\it I}$\overline{4}${\it m}2 & P:$\langle 4 \rangle$ & W$\Delta$UY:ii;{\bf 2a},ij;{\bf 1a} $\Sigma$F:ii;{\bf 1sa},ij;{\bf 1$\bm{\ell}$} Q:{\bf 1sp}\\
{\bf 120} & {\it I}$\overline{4}${\it c}2 &  & W$\Delta$UY:ii;{\bf 2a},ij;{\bf 1a} $\Sigma$F:ii;{\bf 1sa},ij;{\bf 1$\bm{\ell}$}\\
{\bf 121} & {\it I}$\overline{4}$2{\it m} &  & $\Sigma$F:ii;{\bf 2a},ij;{\bf 1a} Q:ij;{\bf 1a} $\Delta$UY:ii;{\bf 1sa},ij;{\bf 1$\bm{\ell}$}\\
{\bf 122} & {\it I}$\overline{4}$2{\it d} & P:$\langle 5 \rangle$ & W:$\langle 3 \rangle$ $\Sigma$F:ii;{\bf 2a},ij;{\bf 1a} $\Delta$:ii;{\bf 1sa},ij;{\bf 1$\bm{\ell}$} Q:ij;{\bf 1a}\\
\end{tabular}
}
\caption{\label{tab:S3}Patterns of gap-closing points after parametric gap closing for tetragonal space groups.}
\end{table}

\begin{table}[h]
\centering
{\footnotesize
\begin{tabular}{cc|c|c} 
\multicolumn{2}{c|}{space group}& point & line\\
\hline
{\bf 143} & {\it P}3 &  & $\Delta$P:ij;{\bf 1a}\\
{\bf 144} & {\it P}$3_{1}$ &  & $\Delta$P:ij;{\bf 1a}\\
{\bf 145} & {\it P}$3_{2}$ &  & $\Delta$P:ij;{\bf 1a}\\
{\bf 146} & {\it R}3 &  & $\Lambda$P:ij;{\bf 1a}\\
{\bf 149} & {\it P}312 &  & $\Delta$P:ii;{\bf 3a},ij;{\bf 1a} UTST'S':{\bf 1p} $\Sigma$R:ij;{\bf 1a}\\
{\bf 150} & {\it P}321 & KH:(3,4);{\bf 3ca} & $\Delta$:ii;{\bf 3a},ij;{\bf 1a} U$\Sigma$R:{\bf 1p} PTST'S':ij;{\bf 1a}\\
{\bf 151} & {\it P}$3_{1}$12 &  & $\Delta$P:ii;{\bf 3a},ij;{\bf 1a} UTST'S':{\bf 1p} $\Sigma$R:ij;{\bf 1a}\\
{\bf 152} & {\it P}$3_{1}$21 & KH:(3,4);{\bf 3ca} & $\Delta$:ii;{\bf 3a},ij;{\bf 1a} U$\Sigma$R:{\bf 1p} PTST'S':ij;{\bf 1a}\\
{\bf 153} & {\it P}$3_{2}$12 &  & $\Delta$P:ii;{\bf 3a},ij;{\bf 1a} UTST'S':{\bf 1p} $\Sigma$R:ij;{\bf 1a}\\
{\bf 154} & {\it P}$3_{2}$21 & KH:(3,4);{\bf 3ca} & $\Delta$:ii;{\bf 3a},ij;{\bf 1a} U$\Sigma$R:{\bf 1p} PTST'S':ij;{\bf 1a}\\
{\bf 155} & {\it R}32 &  & $\Lambda$P:ij;{\bf 3a},ij;{\bf 1a} B$\Sigma$QY:ij;{\bf 1a}\\
{\bf 156} & {\it P}3{\it m}1 & KH:ij;{\bf 1a} & $\Delta$:(3,4);{\bf 3$\bm{\ell}$} U$\Sigma$R:ij;{\bf 1$\bm{\ell}$} P:ij;{\bf 1a} TST'S':{\bf 1sp}\\
{\bf 157} & {\it P}31{\it m} &  & $\Delta$P:(3,4);{\bf 3$\bm{\ell}$} UTST'S':ij;{\bf 1$\bm{\ell}$} $\Sigma$R:{\bf 1sp}\\
{\bf 158} & {\it P}3{\it c}1 & K:ij;{\bf 1a} & $\Delta$:(3,4);{\bf 3$\bm{\ell}$} P:ij;{\bf 1a} U$\Sigma$R:ij;{\bf 1$\bm{\ell}$} TT':{\bf 1sp}\\
{\bf 159} & {\it P}31{\it c} &  & $\Delta$P:(3,4);{\bf 3$\bm{\ell}$} UTST'S':ij;{\bf 1$\bm{\ell}$} $\Sigma$:{\bf 1sp}\\
{\bf 160} & {\it R}3{\it m} &  & $\Lambda$P:(3,4);{\bf 3$\bm{\ell}$} B$\Sigma$QY:{\bf 1sp}\\
{\bf 161} & {\it R}3{\it c} &  & $\Lambda$P:(3,4);{\bf 3$\bm{\ell}$} $\Sigma$Q:{\bf 1sp}\\
{\bf 168} & {\it P}6 &  & $\Delta$UP:ij;{\bf 1a} TST'S'$\Sigma$R:{\bf 1p}\\
{\bf 169} & {\it P}$6_{1}$ &  & $\Delta$UP:ij;{\bf 1a} TT'$\Sigma$:{\bf 1p}\\
{\bf 170} & {\it P}$6_{5}$ &  & $\Delta$UP:ij;{\bf 1a} TT'$\Sigma$:{\bf 1p}\\
{\bf 171} & {\it P}$6_{2}$ &  & $\Delta$UP:ij;{\bf 1a} TST'S'$\Sigma$R:{\bf 1p}\\
{\bf 172} & {\it P}$6_{4}$ &  & $\Delta$UP:ij;{\bf 1a} TST'S'$\Sigma$R:{\bf 1p}\\
{\bf 173} & {\it P}$6_{3}$ &  & $\Delta$UP:ij;{\bf 1a} TT'$\Sigma$;{\bf 1p}\\
{\bf 174} & {\it P}$\overline{6}$ & KH:$\langle 6 \rangle$ & $\Delta$:(4,4):{\bf 3scp} U:{\bf 1sp} P:ij;{\bf 1a} TST'S'$\Sigma$R:ij;{\bf 1$\bm{\ell}$}\\
{\bf 177} & {\it P}622 & KH:(3,4);{\bf 3ca} & $\Delta$:ii;{\bf 6a},ij;{\bf 1a} UTST'S'$\Sigma$R:ii;{\bf 2a},ij;{\bf 1a} P:ii;{\bf 3a},ij;{\bf 1a}\\
{\bf 178} & {\it P}$6_{1}$22 & K:(3,4);{\bf 3ca} & $\Delta$:ii;{\bf 6a},ij;{\bf 1a} UTT'$\Sigma$:ii;{\bf 2a},ij;{\bf 1a} P:ii;{\bf 3a},ij;{\bf 1a}\\
{\bf 179} & {\it P}$6_{5}$22 & K:(3,4);{\bf 3ca} & $\Delta$:ii;{\bf 6a},ij;{\bf 1a} UTT'$\Sigma$:ii;{\bf 2a},ij;{\bf 1a} P:ii;{\bf 3a},ij;{\bf 1a}\\
{\bf 180} & {\it P}$6_{2}$22 & KH:(3,4);{\bf 3ca} & $\Delta$:ii;{\bf 6a},ij;{\bf 1a} UTST'S'$\Sigma$R:ii;{\bf 2a},ij;{\bf 1a} P:ii;{\bf 3a},ij;{\bf 1a}\\
{\bf 181} & {\it P}$6_{4}$22 & KH:(3,4);{\bf 3ca} & $\Delta$:ii;{\bf 6a},ij;{\bf 1a} UTST'S'$\Sigma$R:ii;{\bf 2a},ij;{\bf 1a} P:ii;{\bf 3a},ij;{\bf 1a}\\
{\bf 182} & {\it P}$6_{3}$22 & K:(3,4);{\bf 3ca} & $\Delta$:ii;{\bf 6a},ij;{\bf 1a} UTT'$\Sigma$:ii;{\bf 2a},ij;{\bf 1a} P:ii;{\bf 3a},ij;{\bf 1a}\\
{\bf 183} & {\it P}6{\it mm} & KH:(3,4);{\bf 3$\bm{\ell}$} & P:(3,4);{\bf 3$\bm{\ell}$} TST'S'$\Sigma$R:ii;{\bf 1sa},ij;{\bf 1$\bm{\ell}$}\\
{\bf 184} & {\it P}6{\it cc} & K:(3,4);{\bf 3$\bm{\ell}$} & P:(3,4);{\bf 3$\bm{\ell}$} TT'$\Sigma$:ii;{\bf 1sa},ij;{\bf 1$\bm{\ell}$}\\
{\bf 185} & {\it P}$6_{3}${\it cm} & K:(3,4);{\bf 3$\bm{\ell}$} & P:(3,4);{\bf 3$\bm{\ell}$} TT'$\Sigma$:ii;{\bf 1sa},ij;{\bf 1$\bm{\ell}$}\\
{\bf 186} & {\it P}$6_{3}${\it mc} & K:(3,4);{\bf 3$\bm{\ell}$} & P:(3,4);{\bf 3$\bm{\ell}$} TT'$\Sigma$:ii;{\bf 1sa},ij;{\bf 1$\bm{\ell}$}\\
{\bf 187} & {\it P}$\overline{6}${\it m}2 & KH:$\langle 6 \rangle$ & $\Delta$:(3,4);{\bf 3$\bm{\ell}$} (3,3)(4,4);{\bf 3sca} UTST'S':ii;{\bf 1sa},ij;{\bf 1$\bm{\ell}$} P:ii;{\bf 3a},ij;{\bf 1a}\\
{\bf 188} & {\it P}$\overline{6}${\it c}2 & K:$\langle 6 \rangle$ & $\Delta$:(3,4);{\bf 3$\bm{\ell}$} (3,3)(4,4);{\bf 3sca} UTT':ii;{\bf 1sa},ij;{\bf 1$\bm{\ell}$} P:ii;{\bf 3a},ij;{\bf 1a} R:$\langle 1 \rangle$\\
{\bf 189} & {\it P}$\overline{6}$2{\it m} &  & $\Delta$:(3,4);{\bf 3$\bm{\ell}$} (3,3)(4,4);{\bf 3sca} U$\Sigma$R:ii;{\bf 1sa},ij;{\bf 1$\bm{\ell}$} P:(3,4);{\bf 3$\bm{\ell}$}\\
{\bf 190} & {\it P}$\overline{6}$2{\it c} & H:$\langle 8 \rangle$ & $\Delta$:(3,4);{\bf 3$\bm{\ell}$} (3,3)(4,4);{\bf 3sca} U$\Sigma$:ii;{\bf 1sa},ij;{\bf 1$\bm{\ell}$} P:(3,4);{\bf 3$\bm{\ell}$} SS':$\langle 1 \rangle$\\
\end{tabular}
}
\caption{\label{tab:S4}Patterns of gap-closing points after parametric gap closing for  trigonal and hexagonal space groups. 
}
\end{table}

\begin{table}[h]
\centering
{\footnotesize
\begin{tabular}{cc|c|c} 
\multicolumn{2}{c|}{space group}& point & line\\
\hline
{\bf 195} & {\it P}23 &  & $\Delta$ZT:ii;{\bf 2a},ij;{\bf 1a} $\Sigma$S:{\bf 1p} $\Lambda$:ij;{\bf 1a}\\
{\bf 196} & {\it F}23 & W:ij;{\bf 1a} & $\Delta$Z:ii;{\bf 2a},ij;{\bf 1a} $\Lambda$:ij;{\bf 1a} $\Sigma$S:{\bf 1p}\\
{\bf 197} & {\it I}23 &  & $\Sigma$G:{\bf 1p} $\Delta$:ii;{\bf 2a},ij;{\bf 1a} $\Lambda$DF:ij;{\bf 1a}\\
{\bf 198} & {\it P}$2_{1}$3 &  & $\Delta$:ii;{\bf 2a},ij;{\bf 1a} $\Sigma$:{\bf 1p} $\Lambda$:ij;{\bf 1a}\\
{\bf 199} & {\it I}$2_{1}$3 & P:$\langle 7 \rangle$ & $\Sigma$G:{\bf 1p} $\Delta$:ii;{\bf 2a},ij;{\bf 1a} $\Lambda$DF:ij;{\bf 1a}\\
{\bf 207} & {\it P}432 &  & $\Delta$T:ii;{\bf 4a},ij;{\bf 1a} $\Sigma$SZ:ii;{\bf 2a},ij;{\bf 1a} $\Lambda$:ii;{\bf 3a},ij;{\bf 1a}\\
{\bf 208} & {\it P}$4_{2}$32 &  & $\Delta$T:ii;{\bf 4a},ij;{\bf 1a} $\Sigma$SZ:ii;{\bf 2a},ij;{\bf 1a} $\Lambda$:ii;{\bf 3a},ij;{\bf 1a}\\
{\bf 209} & {\it F}432 &  & $\Delta$:ii;{\bf 4a},ij;{\bf 1a} $\Lambda$:ii;{\bf 3a},ij;{\bf 1a} $\Sigma$SZ:ii;{\bf 2a},ij;{\bf 1a} Q:ij;{\bf 1a}\\
{\bf 210} & {\it F}$4_{1}$32 & W:(2,4);{\bf 4a} & $\Delta$:ii;{\bf 4a},ij;{\bf 1a} $\Lambda$:ii;{\bf 3a},ij;{\bf 1a} $\Sigma$SZ:ii;{\bf 2a},ij;{\bf 1a} Q:ij;{\bf 1a}\\
{\bf 211} & {\it I}432 &  & $\Sigma$DG:ii;{\bf 2a},ij;{\bf 1a} $\Delta$:ii;{\bf 4a},ij;{\bf 1a} $\Lambda$F:ii;{\bf 3a},ij;{\bf 1a}\\
{\bf 212} & {\it P}$4_{3}$32 &  & $\Delta$:ii;{\bf 4a},ij;{\bf 1a} $\Sigma$:ii;{\bf 2a},ij;{\bf 1a} $\Lambda$:ii;{\bf 3a},ij;{\bf 1a}\\
{\bf 213} & {\it P}$4_{1}$32 &  & $\Delta$:ii;{\bf 4a},ij;{\bf 1a} $\Sigma$:ii;{\bf 2a},ij;{\bf 1a} $\Lambda$:ii;{\bf 3a},ij;{\bf 1a}\\
{\bf 214} & {\it I}$4_{1}$32 & P:$\langle 7 \rangle$ & $\Sigma$DG:ii;{\bf 2a},ij;{\bf 1a} $\Delta$:ii;{\bf 4a},ij;{\bf 1a} $\Lambda$F:ii;{\bf 3a},ij;{\bf 1a}\\
{\bf 215} & {\it P}$\overline{4}$3{\it m} &  & $\Sigma$S:ii;{\bf 1sa},ij;{\bf 1$\bm{\ell}$} $\Lambda$:(3,4);{\bf 3$\bm{\ell}$} Z:ii;{\bf 2a},ij;{\bf 1a}\\
{\bf 216} & {\it F}$\overline{4}$3{\it m} & W:$\langle 4 \rangle$ & $\Lambda$:(3,4);{\bf 3$\bm{\ell}$} $\Sigma$S:ii;{\bf 1sa},ij;{\bf 1$\bm{\ell}$} Z:ii;{\bf 2a},ij;{\bf 1a} Q:{\bf 1sp}\\
{\bf 217} & {\it I}$\overline{4}$3{\it m} &  & $\Sigma$G:ii;{\bf 1sa},ij;{\bf 1$\bm{\ell}$} $\Lambda$(3,4);{\bf 3$\bm{\ell}$}\\
{\bf 218} & {\it P}$\overline{4}$3{\it n} &  & $\Sigma$:ii;{\bf 1sa},ij;{\bf 1$\bm{\ell}$} $\Lambda$:(3,4);{\bf 3$\bm{\ell}$} Z:ii;{\bf 2a},ij;{\bf 1a}\\
{\bf 219} & {\it F}$\overline{4}$3{\it c} &  & $\Lambda$:(3,4);{\bf 3$\bm{\ell}$} $\Sigma$S:ii;{\bf 1sp},ij;{\bf 1$\bm{\ell}$} Z:ii;{\bf 2a},ij;{\bf 1a}\\
{\bf 220} & {\it I}$\overline{4}$3{\it d} & P:(17,18);{\bf 6$\bm{\ell}$} & $\Sigma$:ii;{\bf 1sa},ij;{\bf 1$\bm{\ell}$} $\Lambda$:(3,4);{\bf 3$\bm{\ell}$} D:$\langle 1 \rangle$ F:(9,10);{\bf 3$\bm{\ell}$}\\
\end{tabular}
}
\caption{\label{tab:S5}Patterns of gap-closing points after parametric gap closing for cubic space groups.
}
\end{table}


\begin{thebibliography}{99}
\bibitem{P_LuSI}
Because of the extremely small energy gap, 
numerical error in the structural optimization is larger than the energy scale around the Weyl point
at ambient pressure,
Therefore, in our calculation, we simply multiply lattice constant $c$ by $0.945$ instead of structural optimization.
A realistic value of the transition pressure to the Weyl semimetal phase is an open question.
\end{thebibliography}
\end{document}